\documentclass[12pt]{article} %,a4paper

\usepackage{physics}
\usepackage{graphicx}
\usepackage{amsmath}
\usepackage{amsthm}
\usepackage{amssymb}
\usepackage{hyperref}
\usepackage[paperwidth=8.5in,paperheight=11in,textwidth=6.25in,textheight=9.1in,
 centering]{geometry}
\usepackage{tikz}
\usepackage{cite}
\usepackage{amscd}
\usepackage{setspace}
\usepackage{bbm}
\usepackage{dsfont}
\usepackage{cancel}
\usepackage{relsize}
\usepackage[utf8]{inputenc}
\usepackage{mathtools}
\usepackage{newtxtext,newtxmath}
\numberwithin{equation}{section}

\hypersetup{
  pdftitle={Conical Defects in JT Gravity from BF Theory: Quantization, Fusion, and Weighted Moduli Spaces},
  pdfauthor={Wei Gu}
}

\begin{document}
\begin{titlepage}

\begin{flushright}

\end{flushright}

\vskip 3cm

\begin{center}
{\Large \bf
Conical Defects in JT Gravity from BF Theory:\\[0.3em]
Quantization, Fusion, and Weighted Moduli Spaces}
\vskip 2.0cm

 Wei Gu \\

\bigskip
\begin{tabular}{cc}
Zhejiang Institute of Modern Physics, School of Physics,  Zhejiang University\\
Hangzhou, Zhejiang 310058, China\\

 \end{tabular}

\vskip 1cm

\textbf{Abstract}
\end{center}

We construct a gauge-invariant BF representative of a conical defect in
Jackiw--Teitelboim gravity, with its elliptic direction fixed by the
reconstruction of the gravitational variables. In the gravitational
sector, the BF observable reproduces the metric defect insertion and its
distributional curvature source, and fixes the holonomy around the defect
to an elliptic conjugacy class. For a single defect on the disk, the
position integral combines with the residual
\(\mathrm{PSL}(2,\mathbb R)\) quotient, leaving a fixed elliptic sector.
The exact boundary quantization of this sector gives the elementary kernel
and one-defect disk amplitude directly, without using a matrix model. The
same local BF operator describes both sharp and blunt defects. Their
difference appears only when several defect positions are integrated. BF
source composition makes the fractional deficits additive. Requiring a
fused source to remain in the gravitational cone sector gives the
admissibility condition for defect collisions, while positivity of the
hyperbolic area on each component gives the corresponding stability
condition. These conditions coincide with those for Hassett weighted
curves. Combining these BF results with the known Hassett compactification
and conical Weil--Petersson geometry yields the fixed-order genus-zero
contact terms and recovers the cluster expansion of deformed JT gravity.
Thus sharp and blunt defects differ through the compactified geometry of
their relative positions, rather than through the elementary BF operator.

\medskip
\noindent

\bigskip
\vfill
\end{titlepage}

\setcounter{tocdepth}{2}
\tableofcontents

\section{Introduction}
\label{sec:introduction}

JT gravity~\cite{Jackiw:1984je,Teitelboim:1983ux} provides one of the simplest settings in which we can study questions in quantum gravity explicitly. This setting can be extended by including conical defects~\cite{Mertens:2019tcm}, while much of its exact solvability is still preserved. In particular, exponential deformations of the JT dilaton potential can be expanded in terms of integrated conical defects and have a corresponding matrix-model description~\cite{Witten:2020wvy,Maxfield:2020ale}. An important subtlety arises at higher orders in this expansion. When the defects are sufficiently blunt, several defects are allowed to collide, and the corresponding amplitudes receive generalized Weil--Petersson contact terms~\cite{Eberhardt:2023rzz}. These collisions are naturally described by weighted compactifications~\cite{Hassett:2003,Anagnostou:2022cmq} and the associated wall-crossing relations~\cite{Anagnostou:2023jiy}. We refer to~\cite{Dijkgraaf:2018vnm} and~\cite{Mertens:2022irh} for reviews of topological gravity and intersection theory, and of Jackiw--Teitelboim gravity, respectively.

Following~\cite{Witten:2020wvy}, we expand exponential deformations of the JT dilaton potential in integrated spacetime defects. A defect species is specified by its fractional deficit \(q\) and its fugacity. We formulate
this expansion directly in BF theory. We first construct and quantize the elementary defect operator, and then study the composition of several integrated sources and determine which of their collisions remain in the gravitational sector.

A complementary motivation comes from two-dimensional BF theory with a compact gauge group. After integrating out the adjoint scalar, the gauge connection is constrained to be flat. Thus, after quotienting by gauge transformations, the path integral is supported on the moduli space of flat connections, whose natural symplectic structure is the Atiyah--Bott--Goldman symplectic form~\cite{Atiyah:1983yg,Goldman:1984aa}. More generally, gauge theory is related to intersection-theoretic structures in various settings. One well-known example is the relation between the Verlinde algebra and
the quantum cohomology of the Grassmannian, for which Witten gave a
quantum-field-theoretic explanation~\cite{Witten:1993xi}. Similarly, correspondences between topological quantum field theories can lead to relations between their associated quantum cohomology theories~\cite{Gu:2025gtb}.

Ordinary JT gravity admits a first-order formulation as an
$\mathrm{SL}(2,\mathbb{R})$ BF theory~\cite{Blommaert:2018oue}. Its quantization
gives exact partition functions, Wilson-line observables, and boundary
correlation functions~\cite{Iliesiu:2019xuh}. It is then natural to ask how
much of the geometry of the defect-position moduli can be understood
directly in BF theory. In particular, we study whether the cotangent-line
bundle associated with a moving defect, as well as the conditions governing
the fusion and collision of several defects, can be described in terms of
BF observables and localization. We further ask how these structures are
related to the known weighted compactification of the defect-position
space.\footnote{
The original motivation for this work came from a question that the author
posed to E.~Witten: whether the $\psi$-classes on the moduli space of
Riemann surfaces admit a direct realization in noncompact BF theory, for
example through observables of the form
$\operatorname{Tr}e^{i\alpha X}$. Witten pointed out that related aspects had been studied in his work on
deformations of JT gravity and in related developments
\cite{Witten:2020wvy,Maxfield:2020ale,Eberhardt:2023rzz}. These references led us to the spacetime-defect
formulation and eventually to the BF description developed in this paper.
}

Previous work provides the geometric and gravitational ingredients needed
to address this question. Exponential deformations of the JT dilaton
potential can be expanded in terms of integrated conical defects, and the
corresponding matrix-model description was first developed in the sharp
regime~\cite{Witten:2020wvy,Maxfield:2020ale}. The same angular range also
appears in mathematical results on hyperbolic cone surfaces
\cite{Tan:2004pb}. The matrix-model description was extended to arbitrary
opening angles through the deformed minimal string, where the nonlinear
cluster terms of the blunt regime were found explicitly
\cite{Turiaci:2020fjj}.

The moduli spaces of hyperbolic surfaces with conical defects carry
angle-dependent Weil--Petersson classes, together with additional contact
terms supported on the admissible collision loci~\cite{Eberhardt:2023rzz}.
Their natural compactifications are the weighted Hassett
spaces~\cite{Hassett:2003,Anagnostou:2022cmq}. As the defect weights vary,
the markings that are allowed to coincide also change, leading to
wall-crossing relations for the corresponding volume
polynomials~\cite{Anagnostou:2023jiy}. This global geometric description
leaves an important gauge-theoretic question open: are sharp and blunt
defects described by different local BF operators, or do they share the
same elementary observable and differ only through the chamber-dependent
compactification of their relative-position moduli?

In this work, we provide a common BF description of sharp and blunt
defects. After choosing the elliptic reduction that defines the
gravitational sector, we construct a gauge-invariant BF representative of
the integrated metric defect and show how it enters the gravitational path
integral. When several defects coincide, we find from the BF source equation
that their fractional deficits are additive. Restricting the fused source
to the gravitational integration cycle then determines whether the
collision remains within the cone sector.

In the open sharp regime, the defect markings are not allowed to coincide,
and their moduli are described by the ordinary Deligne--Mumford
compactification. In the blunt regime, certain subsets of defects may
collide and fuse into another cone point, or into a cusp on the limiting
wall. These coincidence loci are incorporated by the appropriate weighted
Hassett compactification~\cite{Hassett:2003,Anagnostou:2022cmq,
Anagnostou:2023jiy}. Thus, sharp and blunt defects are described by the same
local BF observable. Their difference arises from the collision strata
admitted by the compactified space of their relative positions.

The BF construction also identifies the reduced quantum problem associated
with a single spacetime defect. On the disk, quotienting by the
transformations that move the defect removes its position modulus, and the
integrated observable reduces to the fixed elliptic sector
\(U(1)_{1-q}\). Using the exact BF/Schwarzian quantization of this sector \cite{Mertens:2019tcm} and expressing its defect factor in our JT spectral and sewing conventions, we obtain the elementary kernel and the one-defect disk amplitude without using a matrix model.

Gravitational amplitudes require a further choice of integration
prescription, denoted by $\Gamma_{\mathrm{grav}}$ and referred to below
as the gravitational integration cycle; its precise role and the sense
in which we use this terminology will be explained below.
Although the BF defect correlators can be defined at arbitrary genus, we
carry out the weighted multi-defect construction explicitly only at genus
zero. A systematic extension to higher genus remains open. The local BF
construction also does not select a unique nonperturbative completion of
the gravitational genus expansion.

To fix our conventions, let \(\theta_i\in[0,2\pi]\) denote the opening angle
of the \(i\)-th conical defect. We define its opening fraction \(a_i\) and
fractional deficit \(q_i\) by
\begin{equation}
a_i
\equiv
\frac{\theta_i}{2\pi},
\qquad
q_i
\equiv
1-a_i
=
1-\frac{\theta_i}{2\pi}.
\label{eq:intro_q_definition}
\end{equation}
Thus, \(2\pi q_i\) is the deficit angle. An ordinary cone point has
\(0<q_i<1\), while the limits \(q_i\to0\) and \(q_i\to1\) correspond,
respectively, to a smooth marked point and a cusp.\footnote{
At \(q_i=0\), the local curvature source vanishes, but the integrated
observable does not become the identity. On the gravitational cycle, it
formally reduces to the area insertion, whose classical evaluation on an
asymptotically AdS$_2$ surface requires a boundary subtraction.
Gauss--Bonnet expresses the resulting renormalized area as a topological
contribution together with a finite boundary term involving \(K-1\).
This geometric subtraction should be distinguished from the normalization
of the quantum marked-point operator, which also involves the quotient by
the position modes and the BF sewing convention; see
Section~\ref{subsec:section3-comparison}.
}
In the following, we use \(q_i\) as the primary label of a defect.

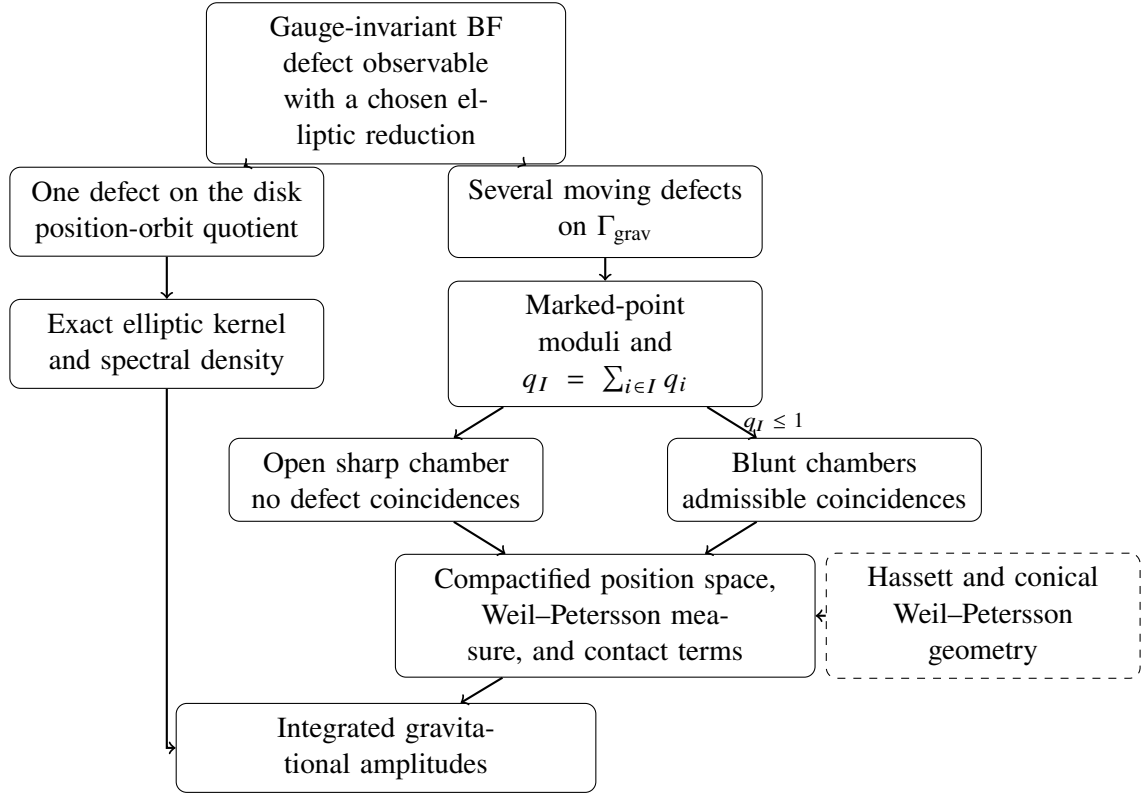
\begin{figure}[t]
\centering

\begin{tikzpicture}[
every node/.style={
    align=center,
    font=\small
},
box/.style={
    draw,
    rounded corners,
    text width=4.4cm,
    minimum height=0.85cm,
    inner sep=5pt
},
smallbox/.style={
    draw,
    rounded corners,
    text width=3.8cm,
    minimum height=0.85cm,
    inner sep=5pt
},
widebox/.style={
    draw,
    rounded corners,
    text width=5.2cm,
    minimum height=0.85cm,
    inner sep=5pt
},
inputbox/.style={
    draw,
    dashed,
    rounded corners,
    text width=3.8cm,
    minimum height=0.85cm,
    inner sep=5pt
}
]

\node[box] (observable) at (0,7.8)
{Gauge-invariant BF defect observable\\
with a chosen elliptic reduction};

\node[smallbox] (onedisk) at (-2.9,6.1)
{One defect on the disk\\
position-orbit quotient};

\node[smallbox] (multidefect) at (2.9,6.1)
{Several moving defects\\
on \(\Gamma_{\mathrm{grav}}\)};

\draw[->,thick] (observable) -- (onedisk);
\draw[->,thick] (observable) -- (multidefect);

\node[smallbox] (kernel) at (-2.9,4.35)
{Exact elliptic kernel\\
and spectral density};

\node[smallbox] (fusion) at (2.9,4.35)
{Marked-point moduli and\\
\(q_I=\sum_{i\in I}q_i\)};

\draw[->,thick] (onedisk) -- (kernel);
\draw[->,thick] (multidefect) -- (fusion);

\node[smallbox] (sharp) at (0.0,2.55)
{Open sharp chamber\\
no defect coincidences};

\node[smallbox] (blunt) at (5.8,2.55)
{Blunt chambers\\
admissible coincidences};

\draw[->,thick] (fusion) -- (sharp);
\draw[->,thick] (fusion) -- node[right, font=\scriptsize]
{\(q_I\leq1\)} (blunt);

\node[widebox] (geometry) at (2.9,0.75)
{Compactified position space,\\
Weil--Petersson measure, and contact terms};

\draw[->,thick] (sharp) -- (geometry);
\draw[->,thick] (blunt) -- (geometry);

\node[inputbox] (input) at (7.9,0.75)
{Hassett and conical\\
Weil--Petersson geometry};

\draw[->,thick,dashed] (input) -- (geometry);

\node[widebox] (amplitudes) at (0,-1.0)
{Integrated gravitational amplitudes};

\draw[->,thick] (geometry) -- (amplitudes);
\draw[->,thick] (kernel) |- (amplitudes);

\end{tikzpicture}

\caption{
The elementary kernel and the multi-defect localization geometry arise from
two reductions of the same BF defect observable. For one defect on the disk,
the position quotient gives the fixed elliptic sector. With several moving
defects, BF source composition determines the admissible collision channels,
while the Hassett compactification and conical Weil--Petersson geometry
supply the global measure and contact terms.
}
\label{fig:overview}
\end{figure}

The main results of this paper are as follows:
\begin{enumerate}

\item
\textbf{A gauge-invariant BF defect observable.}

We introduce a nonpropagating adjoint section \(n\) of fixed norm, which
transforms as \(n\mapsto gng^{-1}\). The section \(n\) represents the
elliptic reduction used to define the gravitational sector. It replaces a
gauge-dependent choice of elliptic generator with covariantly transforming
background data. Relative to this reduction, we define
\begin{equation}
\mathfrak D_q^{\mathrm{BF}}[n]
=
\int_\Sigma
\Omega_n(A)\,
\exp\!\left(
2\pi iq\langle n,X\rangle
\right).
\label{eq:intro_integrated}
\end{equation}
Both \(\Omega_n(A)\) and \(\langle n,X\rangle\) are invariant under
simultaneous gauge transformations of \(A\), \(X\), and \(n\).

On the gravitational integration cycle, we may locally choose \(n=J\), for
which
\begin{equation}
\langle J,X\rangle=i\Phi,
\qquad
\Omega_J(A)=\sqrt g\,\mathrm d^2x.
\label{eq:intro_grav_gauge}
\end{equation}
It then follows that
\begin{equation}
\left.
\mathfrak D_q^{\mathrm{BF}}
\right|_{\Gamma_{\mathrm{grav}}}
=
\int_\Sigma
\sqrt g\,\mathrm d^2x\,
e^{-2\pi q\Phi}.
\label{eq:intro_JT_defect}
\end{equation}
Thus, the BF observable reproduces the standard integrated metric defect,
including its position measure. Correlators of these observables are defined
by restricting the BF path integral to \(\Gamma_{\mathrm{grav}}\), and they
admit the usual cutting-and-sewing representation of BF
theory~\cite{Iliesiu:2019xuh}.

\item
\textbf{Localization and marked-point geometry.}

Integrating over \(X\) imposes flatness away from the insertions and fixes an
elliptic holonomy around each defect~\cite{Blommaert:2018oue,
Iliesiu:2019xuh}. On \(\Gamma_{\mathrm{grav}}\), the localized
configurations are hyperbolic surfaces with marked conical points.
Therefore, the position of an integrated BF defect becomes a marked point
on the corresponding cone surface.

After quotienting by the two transformations that move the marked point,
the remaining stabilizer is the compact rotation subgroup
\(SO(2)\simeq U(1)\). As the marked point and the localized
surface vary, the local frames at the marked point form a principal
\(U(1)\) bundle. Its weight-\(-1\) representation gives the
cotangent-line bundle
\begin{equation}
\mathcal L_i=T_{p_i}^{*}\Sigma,
\qquad
\psi_i=c_1(\mathcal L_i).
\label{eq:intro_psi}
\end{equation}
Thus, the cotangent-line class records the twisting of the residual
rotational frame over the moduli space of the moving defect.

\item
\textbf{The elementary quantum defect kernel.}

For a single defect on the disk, quotienting by its position modes reduces
the observable to the elliptic sector with opening angle \(2\pi(1-q)\).
Using the exact BF/Schwarzian quantization of this
sector~\cite{Mertens:2019tcm}, we obtain
\begin{equation}
Z_{1\text{-}\mathrm{def}}(\beta;q)
=
\int_0^\infty
\mathrm dE\,
\frac{
\cosh\!\left(2\pi(1-q)\sqrt E\right)
}{
2\pi\sqrt E
}\,
e^{-\beta E}.
\label{eq:intro_spectral}
\end{equation}
This result agrees with the elliptic continuation of the exact JT trumpet
and with the known one-defect amplitude~\cite{Saad:2019lba,
Witten:2020wvy,Maxfield:2020ale}. The same elementary kernel applies
throughout the elliptic range. Therefore, sharp and blunt defects require
neither different local BF operators nor different local quantizations.
Their distinction first appears when collisions among several integrated
defects are considered.

\item
\textbf{Fusion and collision admissibility.}

When the defects in a subset \(I\) approach a common point, their aligned BF
sources combine according to
\begin{equation}
q_I=\sum_{i\in I}q_i.
\label{eq:intro_fusion}
\end{equation}
Requiring the fused source to remain in the compactified gravitational cone
sector gives
\begin{equation}
q_I\leq1.
\label{eq:intro_weight}
\end{equation}
For \(q_I<1\), the fused source is another conical point, while
\(q_I=1\) gives the cuspidal limit. The same inequality is the
coincidence condition for marked points with Hassett weights \(q_i\).
The stability condition also follows from the gravitational restriction.
Applying conical Gauss--Bonnet to each irreducible component and requiring
its hyperbolic area to be positive gives
\begin{equation}
2g_v-2+c_v+\sum_{i\in I_v}q_i
>
0,
\label{eq:component-weighted-stability}
\end{equation}
where \(c_v\) counts the nodes and geodesic boundaries on the component.
This is precisely the weighted stability condition for the corresponding
component of a Hassett stable curve. Thus the BF source-composition rule,
together with the restriction to the gravitational sector, reproduces
both the local coincidence condition and the componentwise stability
condition underlying the weighted compactification. The global
compactification, its reduction morphisms, and the coefficients associated
with the collision strata are provided by the known Hassett and conical
Weil--Petersson geometry
\cite{Hassett:2003,Anagnostou:2022cmq,
Anagnostou:2023jiy,Eberhardt:2023rzz}.

The endpoint \(q=0\) of the BF family describes a smooth moving marked
point with vanishing localized curvature source. Its exact disk matrix
element fixes
\begin{equation}
Z_{\mathfrak D_0}^{\mathrm{disk}}(\beta)
=2\beta Z_{\mathrm{JT}}^{\mathrm{disk}}(\beta).
\label{eq:intro-source-free-normalization}
\end{equation}

while the first variation at this endpoint gives the corresponding dilaton
insertion. Together with the ordinary or weighted forgetful-map identities,
these BF identifications reproduce the string and dilaton relations, with
their normalization fixed by the elementary kernel. The same kernel,
combined with the compactified moduli-space data, determines the integrated
defect amplitudes summarized in Fig.~\ref{fig:overview}.

\end{enumerate}

We use ``integration cycle'' in the standard formal sense of a contour in
the complexified BF field space on which the path integral is defined.
The BF action alone does not determine this contour, since the integration
prescription provides additional global data.\footnote{
More generally, an integration cycle may be part of the definition of a
quantum theory rather than a consequence of its local action. This
viewpoint is familiar from Picard--Lefschetz theory, analytically continued
path integrals, and Chern--Simons theory
\cite{Witten:2010zr,Witten:2010cx}.
}
Throughout this paper, \(\Gamma_{\mathrm{grav}}\) denotes the gravitational
choice of this contour. It is characterized by the condition that the BF
variables reconstruct real, nondegenerate, orientation-preserving JT
geometries, together with the appropriate contour for the BF scalar \(X\).
We use this physical characterization and do not attempt a global
construction of \(\Gamma_{\mathrm{grav}}\) in the complexified field
space. On \(\Gamma_{\mathrm{grav}}\), the exact elliptic kernel fixes the
one-defect quantum data. For fused sources, the same gravitational
restriction determines whether a collision remains in the cone sector.
Once a collision is admissible, the compactification of the
relative-position moduli determines how its stratum contributes to the
multi-defect amplitude.

The gauge-invariant BF correlators can be defined at any fixed topology,
although the weighted multi-defect construction is evaluated explicitly
here only at genus zero. At the smooth-marking endpoint of the BF defect
family, the localized curvature source vanishes while the integrated
observable represents an additional smooth moving marked point; its
normalized first variation gives the dilaton insertion. Their disk
normalizations are fixed by the exact elementary kernel. Together with
the corresponding relations on the compactified moduli space, these local
BF operator identifications reproduce the expected string and dilaton
equations. This gives a BF interpretation of the corresponding
insertions and their sewing, but not an independent derivation of the
global forgetful-map and pushforward identities from local BF gauge
symmetry. Deriving these identities directly from BF or BRST Ward
identities remains open, as does the extension of the weighted calculation
and its possible Virasoro structure to higher genus.

The paper is organized as follows. In Section~\ref{sec:BF-defects}, we formulate JT gravity in BF variables, construct the gauge-invariant defect observable, and show that its localized source produces the expected conical geometry. In Section~\ref{sec:quantization}, we reduce the one-defect disk problem to a fixed elliptic sector and use its exact boundary quantization to obtain the elementary kernel and disk amplitude. In Section~\ref{sec:fusion}, we study the composition of localized BF sources, determine when fused sources remain in the gravitational sector, relate this condition to weighted Hassett compactifications, and identify the BF operators entering the string and dilaton relations. In Section~\ref{sec:partition}, we use the elementary BF kernel as the local defect input and the known Weil--Petersson geometry as the global input. In the sharp chamber, the stable amplitudes follow from ordinary Weil--Petersson sewing. In the blunt chambers, the admissible collision strata give the wall-crossing contact terms and reproduce the finite genus-zero cluster expansion.
Finally, Section~\ref{sec:discussion} summarizes our results and briefly comments on the gravitational integration prescription, the relation to available matrix-model results, and open questions concerning the higher-genus weighted construction.

\section{Spacetime defects in the BF formulation of JT gravity}
\label{sec:BF-defects}

We begin by formulating spacetime defects directly in BF theory. We first
review the relation between Euclidean JT gravity and its first-order
formulation as an \(\mathrm{SL}(2,\mathbb{R})\) BF
theory~\cite{Jackiw:1984je,Teitelboim:1983ux,Iliesiu:2019xuh}, paying
particular attention to the gravitational integration
cycle~\cite{Blommaert:2018oue,Iliesiu:2019xuh}. We then choose the elliptic
reduction associated with the gravitational sector and construct, relative
to this reduction, a gauge-invariant BF representative of a conical metric
defect. We show that, at a fixed position, this observable produces the
expected curvature source and elliptic monodromy.

The BF description separates the local data of a defect from the global
geometry of its position moduli. Locally, the insertion determines the
curvature source and the corresponding elliptic conjugacy class. Conical
defects are conventionally divided into sharp and blunt regimes according
to their opening angles~\cite{Maxfield:2020ale,Turiaci:2020fjj,
Eberhardt:2023rzz}. We find, however, that this distinction does not require
different elementary BF observables. It becomes relevant to the
gravitational path integral only when the positions of several defects are
integrated. The gravitational admissibility condition then determines
which collision limits are allowed, while the corresponding compactification
determines which collision strata contribute.

Section~\ref{subsec:jt-and-defects} introduces our geometric conventions.
In Section~\ref{subsec:BF-defect}, we construct the BF defect observable,
and in Section~\ref{subsec:defect-sources}, we derive its source and
composition rules.

\subsection{JT gravity and its BF formulation}
\label{subsec:jt-and-defects}

\paragraph{Metric formulation.}

We work in Euclidean signature and set the AdS radius to one. The JT action
is~\cite{Teitelboim:1983ux,Jackiw:1984je,Saad:2019lba}
\begin{align}
I_{\mathrm{JT}}
={}&
-\frac{S_0}{4\pi}
\left[
\int_{\Sigma}\mathrm{d}^2x\,\sqrt{g}\,R
+
2\int_{\partial\Sigma}\mathrm{d}u\,\sqrt{h}\,K
\right]
\nonumber\\
&
-\frac{1}{2}
\int_{\Sigma}\mathrm{d}^2x\,\sqrt{g}\,
\Phi(R+2)
-
\int_{\partial\Sigma}\mathrm{d}u\,\sqrt{h}\,
\Phi(K-1).
\label{eq:JT-metric-action}
\end{align}
Here \(g_{\mu\nu}\) is the bulk metric, \(h\) is the induced metric on the
asymptotic boundary, \(R\) is the scalar curvature, \(K\) is the extrinsic
curvature, and \(\Phi\) is the dilaton. Variation with respect to \(\Phi\)
gives
\begin{equation}
R+2=0.
\label{eq:JT-curvature-equation}
\end{equation}

\paragraph{BF formulation and gravitational integration cycle.}

To make the sign and normalization conventions explicit, we use the following
real two-dimensional representation of \(\mathfrak{sl}(2,\mathbb{R})\):
\begin{equation}
J
=
\frac12
\begin{pmatrix}
0&1\\
-1&0
\end{pmatrix},
\qquad
P_1
=
\frac12
\begin{pmatrix}
-1&0\\
0&1
\end{pmatrix},
\qquad
P_2
=
\frac12
\begin{pmatrix}
0&1\\
1&0
\end{pmatrix}.
\label{eq:explicit-sl2-basis}
\end{equation}
These generators obey
\begin{equation}
[J,P_1]=P_2,
\qquad
[J,P_2]=-P_1,
\qquad
[P_1,P_2]=-J.
\label{eq:sl2-commutation-relations}
\end{equation}
We use the invariant bilinear form
\begin{equation}
\langle T_1,T_2\rangle
=
2\operatorname{Tr}(T_1T_2),
\label{eq:sl2-bilinear-form}
\end{equation}
for which
\begin{equation}
\langle J,J\rangle=-1,
\qquad
\langle P_a,P_b\rangle=\delta_{ab},
\qquad
\langle J,P_a\rangle=0.
\label{eq:sl2-generator-norms}
\end{equation}

\paragraph{Global form of the gauge group.}
The local BF equations depend only on the Lie algebra
$\mathfrak{sl}(2,\mathbb R)$.  For the gravitational character variety
and its identification with the moduli space of oriented hyperbolic
structures, however, we take the effective gauge group to be
$\mathrm{PSL}(2,\mathbb R)=\mathrm{SL}(2,\mathbb R)/\{\pm\mathbf 1\}$.
We nevertheless use the defining $\mathrm{SL}(2,\mathbb R)$ matrices when
displaying Lie-algebra generators and convenient lifts of holonomies.
These lifts retain additional central information and should not be
identified with the corresponding $\mathrm{PSL}(2,\mathbb R)$ holonomies.
In particular, with our normalization of the compact generator,
\begin{equation}
\exp(2\pi J)=-\mathbf 1
\qquad\text{in }SL(2,\mathbb R),
\end{equation}
whose image is the identity in $\mathrm{PSL}(2,\mathbb R)$.

Following the gravitational convention of~\cite{Saad:2019lba}, we decompose
the connection as
\begin{equation}
A
=
-\omega J+e^aP_a
=
\frac12
\begin{pmatrix}
-e^1 & e^2-\omega\\
e^2+\omega & e^1
\end{pmatrix}.
\label{eq:BF-connection-decomposition}
\end{equation}
Its curvature is
\begin{equation}
F_A
=
\mathrm{d}A+A\wedge A
=
T^aP_a
-
\left(
\mathrm{d}\omega+e^1\wedge e^2
\right)J,
\label{eq:BF-curvature-decomposition}
\end{equation}
where
\begin{equation}
T^1
=
\mathrm{d}e^1+\omega\wedge e^2,
\qquad
T^2
=
\mathrm{d}e^2-\omega\wedge e^1.
\label{eq:torsion-components}
\end{equation}

We similarly decompose the adjoint scalar as
\begin{equation}
X
=
\varphi J+\chi^aP_a
=
\frac12
\begin{pmatrix}
-\chi^1 & \chi^2+\varphi\\
\chi^2-\varphi & \chi^1
\end{pmatrix}.
\label{eq:BF-scalar-decomposition}
\end{equation}
With the bilinear form \eqref{eq:sl2-bilinear-form},
\begin{equation}
\langle J,X\rangle=-\varphi,
\qquad
\langle P_a,X\rangle=\chi_a.
\label{eq:BF-scalar-components}
\end{equation}

The dynamical part of JT gravity is described by the BF action
\cite{Blommaert:2018oue,Iliesiu:2019xuh}
\begin{equation}
I_{\mathrm{BF}}[A,X]
=
-i\int_\Sigma
\langle X,F_A\rangle
+
I_{\partial}^{\mathrm{BF}}[A,X],
\label{eq:JT-BF-action}
\end{equation}
where \(I_{\partial}^{\mathrm{BF}}\) is the boundary term appropriate to the
asymptotically \(\mathrm{AdS}_2\) variational problem. On the gravitational
cycle it reproduces the dilaton boundary term in
\eqref{eq:JT-metric-action}. Since the defect construction below involves
only the bulk fields, we leave this term implicit when it plays no direct
role. The \(S_0\) term provides the usual topological weight and will
likewise be treated separately from the dynamical BF sector.

Using \eqref{eq:BF-curvature-decomposition} and
\eqref{eq:BF-scalar-decomposition}, the bulk action becomes
\begin{equation}
I_{\mathrm{BF}}^{\mathrm{bulk}}
=
-i\int_\Sigma
\left[
\chi_aT^a
+
\varphi
\left(
\mathrm{d}\omega+e^1\wedge e^2
\right)
\right].
\label{eq:BF-action-components}
\end{equation}

The Euclidean gravitational path integral requires a choice of integration
cycle in the complexified BF field space. We denote the gravitational cycle
by \(\Gamma_{\mathrm{grav}}\). For the bulk fields relevant here, it is
characterized by taking \(e^a\) to be real and nondegenerate, with
\begin{equation}
\det(e^a{}_{\mu})>0,
\qquad
g_{\mu\nu}
=
\delta_{ab}e^a{}_{\mu}e^b{}_{\nu},
\label{eq:gravitational-cycle-metric}
\end{equation}
relative to the chosen orientation. The elliptic component of \(X\) is
integrated along the imaginary contour
\begin{equation}
\varphi=-i\Phi,
\qquad
\Phi\in\mathbb{R}.
\label{eq:Euclidean-X-component-contour}
\end{equation}
Since \(\langle J,J\rangle=-1\), this is equivalently
\begin{equation}
\langle J,X\rangle=i\Phi.
\label{eq:Euclidean-X-contour}
\end{equation}
We take the translational components \(\chi^a\) to be real. Their integration
imposes the torsion constraints
\begin{equation}
T^a=0.
\label{eq:BF-torsion-constraint}
\end{equation}

On \(\Gamma_{\mathrm{grav}}\), the remaining bulk term is
\begin{equation}
-i\int_\Sigma
\varphi
\left(
\mathrm{d}\omega+e^1\wedge e^2
\right)
=
-\int_\Sigma
\Phi
\left(
\mathrm{d}\omega+e^1\wedge e^2
\right).
\label{eq:BF-to-metric-dilaton-term}
\end{equation}
For the torsionless connection,
\begin{equation}
\mathrm{d}\omega
=
\frac12R\,e^1\wedge e^2,
\qquad
e^1\wedge e^2
=
\sqrt{g}\,\mathrm{d}^2x,
\label{eq:curvature-volume-relations}
\end{equation}
and therefore
\begin{equation}
I_{\mathrm{BF}}^{\mathrm{bulk}}
\big|_{\Gamma_{\mathrm{grav}}}
=
-\frac12
\int_\Sigma
\mathrm{d}^2x\,\sqrt{g}\,
\Phi(R+2),
\label{eq:BF-metric-bulk-matching}
\end{equation}
after imposing \(T^a=0\). This is precisely the dynamical bulk term of the
metric JT action.

The restriction to \(\Gamma_{\mathrm{grav}}\) is essential. A generic flat $\mathrm{PSL}(2,\mathbb R)$ connection in the full character
variety need not lie in the gravitational component or define a
nondegenerate zweibein, nor does a generic BF configuration admit an interpretation as a
hyperbolic metric
\cite{Blommaert:2018oue,Iliesiu:2019xuh}. The integration cycle is therefore
part of the definition of the Euclidean gravitational path integral. Other
cycles may contain BF sectors without the same geometric interpretation;
they are not included in the gravitational amplitudes considered below.

\paragraph{Metric defects and their semiclassical insertion.}

Exponential deformations of JT gravity can be interpreted as a gas of
conical defects
\cite{Witten:2020wvy,Maxfield:2020ale,Turiaci:2020fjj}. In our conventions,
consider
\begin{equation}
I_{\mathrm{def}}
=
-\lambda
\int_\Sigma
\mathrm{d}^2x\,\sqrt{g}\,
e^{-2\pi q\Phi},
\qquad
0<q<1.
\label{eq:exponential-deformation}
\end{equation}
The corresponding defect has deficit angle \(2\pi q\) and opening angle
\(2\pi(1-q)\). Expanding the path integral to first order in \(\lambda\)
gives
\begin{equation}
Z(\beta;\lambda)
=
Z_{\mathrm{JT}}(\beta)
+
\lambda
\left\langle
\mathfrak{D}^{\mathrm{metric}}_q
\right\rangle_{\mathrm{JT}}
+
\mathcal{O}(\lambda^2),
\label{eq:first-order-defect-expansion}
\end{equation}
with
\begin{equation}
\mathfrak{D}^{\mathrm{metric}}_q
=
\int_\Sigma
\mathrm{d}^2x\,\sqrt{g}\,
e^{-2\pi q\Phi}.
\label{eq:metric-defect}
\end{equation}
At fixed position \(p\), variation with respect to the dilaton gives
\begin{equation}
(R+2)\,\mathrm{vol}_g
=
4\pi q\,\delta_p,
\qquad
\int_\Sigma\delta_p=1,
\label{eq:sourced-curvature-equation}
\end{equation}
where \(\delta_p\) is the distributional two-form Poincar\'e dual to \(p\).
This is the curvature source of a cone point with deficit angle \(2\pi q\).

It is useful to compare the classical evaluation of
\eqref{eq:metric-defect} with the exact one-defect amplitude. On the
undeformed hyperbolic disk saddle
\cite{Maldacena:2016upp,Saad:2019lba,Witten:2020wvy},
\begin{equation}
\mathrm{d}s^2
=
\mathrm{d}\rho^2
+
\sinh^2\rho\,\mathrm{d}\psi^2,
\qquad
\psi\sim\psi+2\pi,
\qquad
\Phi(\rho)
=
\frac{\pi}{\beta}\cosh\rho,
\label{eq:hyperbolic-disk-saddle}
\end{equation}
the integrated insertion evaluates to
\begin{align}
\mathfrak{D}^{\mathrm{cl}}_q
&=
\int_0^{2\pi}\mathrm{d}\psi
\int_0^\infty\mathrm{d}\rho\,
\sinh\rho\,
\exp\!\left[
-\frac{2\pi^2q}{\beta}\cosh\rho
\right]
\nonumber\\
&=
\frac{\beta}{\pi q}
\exp\!\left(
-\frac{2\pi^2q}{\beta}
\right).
\label{eq:semiclassical-defect-integral}
\end{align}
This is only the classical value of the bare integrated insertion on the
undeformed disk. It does not include the quantum measure and fluctuation
factors associated with the conical configuration, nor does the classical
source equation fix the normalization of the corresponding quantum
operator.

The exact answer illustrates both effects. Writing the opening fraction as
\(\alpha=1-q\), the first-order disk result obtained in the matrix-model
definition of the deformed theory
\cite{Witten:2020wvy,Maxfield:2020ale} is, in our conventions,
\begin{equation}
Z(\beta;\lambda)
=
Z_{\mathrm{JT}}(\beta)
\left[
1
+
2\lambda\beta
\exp\!\left(
-\frac{\pi^2q(2-q)}{\beta}
\right)
+
\mathcal{O}(\lambda^2)
\right].
\label{eq:Witten-quantum-result-q}
\end{equation}
Compared with \eqref{eq:semiclassical-defect-integral}, the exponent is
shifted and the factor \(1/(\pi q)\) is absent. The latter cannot be inferred
from the classical curvature source: an angle-dependent rescaling of the
defect operator can always be compensated by the inverse rescaling of its
fugacity. A comparison of prefactors therefore requires a common convention
for both the operator and its coupling.

The same question can also be studied directly in the gauge-fixed
gravitational path integral~\cite{Lin:2023trc}, where the integration over
the defect position carries a nontrivial measure factor. Together with the
fluctuation determinant, this measure reproduces the conical Schwarzian
amplitude. In the BF formulation developed below, we instead define the
defect as a gauge-invariant observable and fix its quantum normalization
using a definite convention for the BF Hilbert space. The matrix-model and
second-order gravitational results will be used only for comparison, rather
than as inputs to our construction.

\subsection{Gauge-invariant spacetime defects in BF theory}
\label{subsec:BF-defect}

Neither \(\sqrt{g}\,\mathrm{d}^2x\) nor \(\Phi\) is defined on the full BF
field space without choosing a gravitational decomposition. We therefore
introduce an elliptic reduction of the adjoint bundle, represented locally
by an adjoint section \(n\) satisfying
\begin{equation}
\langle n,n\rangle
=
\sigma\neq0,
\qquad
n\in\mathcal O_J,
\label{eq:n-fixed}
\end{equation}
where \(\mathcal O_J\) is the adjoint orbit through the elliptic generator
\(J\). Under a gauge transformation,
\begin{equation}
A
\longmapsto
gAg^{-1}+g\,\mathrm{d}g^{-1},
\qquad
X
\longmapsto
gXg^{-1},
\qquad
n
\longmapsto
gng^{-1}.
\label{eq:gauge-transformations-with-n}
\end{equation}
Thus \(n\) replaces the fixed Lie-algebra direction \(J\) by a covariantly
transforming one. Similar fixed-orbit variables appear in coadjoint-orbit
descriptions of non-Abelian charges and Wilson observables
\cite{Alekseev:1988vx,Diakonov:1989un}. Here, however, \(n\) is part of the
background reduction defining the gravitational sector and is not integrated
over in the BF path integral.

Locally one may choose \(n=J\). Globally, such a choice need not be available
in a single gauge: the section \(n\) specifies a reduction to the stabilizer
of the elliptic generator, \(SO(2)\simeq U(1)\). The
construction below depends on this reduction but not on a particular local
representative. In particular, we do not include an independent orbit
integral over \(n\). Such an integral, natural in coadjoint-orbit
constructions of particle and Wilson-line observables
\cite{Iliesiu:2019xuh}, would define a different observable with an
additional orbit measure.

The scalar \(\langle n,X\rangle\) is gauge invariant. To covariantize the
area form, define
\begin{equation}
D_An=\mathrm{d}n+[A,n]
\label{eq:covariant-derivative-n}
\end{equation}
and
\begin{equation}
\Omega_n(A)
=
-\frac{1}{2\sigma}
\left\langle
n,[D_An\wedge D_An]
\right\rangle .
\label{eq:Omega-definition}
\end{equation}
For Lie-algebra-valued one-forms \(u\) and \(v\), the notation
\([u\wedge v]\) denotes the wedge product combined with the Lie bracket.
Since \(D_An\) transforms in the adjoint representation and the bilinear
form is invariant, \(\Omega_n(A)\) is gauge invariant.

To evaluate \(\Omega_n(A)\) on the gravitational cycle, choose locally
\(n=J\). From \eqref{eq:sl2-generator-norms},
\begin{equation}
\sigma
=
\langle J,J\rangle
=
-1.
\label{eq:n-elliptic-norm}
\end{equation}
The commutation relations \eqref{eq:sl2-commutation-relations} give
\begin{equation}
D_AJ
=
[A,J]
=
e^2P_1-e^1P_2,
\label{eq:DAJ}
\end{equation}
and hence
\begin{equation}
[D_AJ\wedge D_AJ]
=
-2e^1\wedge e^2\,J.
\label{eq:DAJ-square}
\end{equation}
Substituting into \eqref{eq:Omega-definition},
\begin{align}
\left.\Omega_n(A)\right|_{n=J}
&=
-\frac{1}{2\sigma}
\left\langle
J,-2e^1\wedge e^2J
\right\rangle
\nonumber\\
&=
e^1\wedge e^2
=
\sqrt{g}\,\mathrm{d}^2x.
\label{eq:Omega-gravitational-limit}
\end{align}
Similarly, using \eqref{eq:Euclidean-X-contour},
\begin{equation}
\left.
\exp\!\left(
2\pi iq\langle n,X\rangle
\right)
\right|_{n=J,\,\Gamma_{\mathrm{grav}}}
=
e^{-2\pi q\Phi}.
\label{eq:exponential-gravitational-limit}
\end{equation}

We therefore define the local defect two-form
\begin{equation}
\mathcal{D}_q(p)
=
\Omega_n(A;p)
\exp\!\left(
2\pi iq\langle n(p),X(p)\rangle
\right)
\label{eq:local-defect}
\end{equation}
and the integrated BF defect
\begin{equation}
\mathfrak{D}^{\mathrm{BF}}_q
=
\int_\Sigma\mathcal{D}_q.
\label{eq:BF-defect}
\end{equation}
More precisely, \(\mathfrak D_q^{\mathrm{BF}}\) is an observable relative to the chosen elliptic reduction. It is invariant under simultaneous gauge transformations of \((A,X,n)\), and its value does not depend on the local representative used to describe that reduction. We suppress this background dependence in the notation. On the gravitational cycle,
\begin{equation}
\left.
\mathfrak{D}^{\mathrm{BF}}_q
\right|_{\Gamma_{\mathrm{grav}}}
=
\int_\Sigma
\sqrt{g}\,\mathrm{d}^2x\,
e^{-2\pi q\Phi}
=
\mathfrak{D}^{\mathrm{metric}}_q.
\label{eq:BF-metric-defect-matching}
\end{equation}
Thus both the dilaton insertion and the measure over the defect position are
recovered from gauge-invariant BF quantities.

For fixed elliptic reductions \(n_i\), the corresponding correlators are
\begin{equation}
\left\langle
\prod_{i=1}^{L}\mathfrak{D}^{\mathrm{BF}}_{q_i}
\right\rangle_{\mathrm{grav}}
=
\int_{\Gamma_{\mathrm{grav}}}
\mathcal{D}A\,\mathcal{D}X\,
e^{-I_{\mathrm{BF}}[A,X]}
\prod_{i=1}^{L}\mathfrak{D}^{\mathrm{BF}}_{q_i}.
\label{eq:defect-correlator}
\end{equation}
Their localization measure on the gravitational component is discussed in
Appendix~\ref{app:moment_map_reduction}.. No additional orbit integral over the \(n_i\) is understood. Gauge-related
triples \((A,X,n)\) and \((A^g,X^g,n^g)\) give the same insertion, since
\(\langle n,X\rangle\) and \(\Omega_n(A)\) are separately gauge invariant.

The gravitational matching does not by itself make
\eqref{eq:Omega-definition} a unique quantum representative. For example,
one may add the gauge-invariant curvature term
\begin{equation}
\Omega_n(A)
\longrightarrow
\Omega_n(A)+c\langle n,F_A\rangle.
\label{eq:Omega-curvature-deformation}
\end{equation}
It vanishes on the smooth flat localization locus but can modify contact
terms when singular configurations or collisions are present. Exact
two-forms and other terms invisible on the smooth locus can lead to similar
boundary contributions. Such ambiguities are familiar in the treatment of
conical defects, particularly in the blunt regime
\cite{Turiaci:2020fjj,Eberhardt:2023rzz,Lin:2023trc,Kruthoff:2024rpx}.

We choose the minimal representative \eqref{eq:Omega-definition}, without
adding any curvature or contact terms. This choice agrees with the metric
operator \eqref{eq:metric-defect}, which contains no independent curvature
coupling. However, the classical matching by itself does not imply that
this representative is unique as a quantum operator. The fusion and
collision rules derived below are therefore understood with respect to this prescription.

\subsection{Defect sources and conical geometry}
\label{subsec:defect-sources}

Consider first a defect inserted at a fixed point \(p\in\Sigma\), with the
integration over its position suppressed. Its \(X\)-dependent part is
\begin{equation}
\mathcal O_q(p)
=
\exp\!\left[
2\pi iq\langle n(p),X(p)\rangle
\right].
\label{eq:fixed-BF-defect}
\end{equation}
Since \(\Omega_n(A)\) is independent of \(X\), it supplies the position
measure without changing the equation imposed by the \(X\) integral. Let
\(\delta_p\) be the Poincar\'e-dual two-form normalized by
\begin{equation}
\int_\Sigma\delta_p=1.
\label{eq:delta-two-form-normalization}
\end{equation}
In the presence of \eqref{eq:fixed-BF-defect}, the \(X\)-dependent exponent
becomes
\begin{equation}
-i\int_\Sigma
\left\langle
X,F_A+2\pi q\,n\,\delta_p
\right\rangle.
\label{eq:combined-source-exponent}
\end{equation}
Integration over \(X\) therefore imposes
\begin{equation}
F_A
=
-2\pi q\,n\,\delta_p.
\label{eq:sourced-BF-equation}
\end{equation}
The connection remains flat away from \(p\).

On \(\Gamma_{\mathrm{grav}}\), choose the local gravitational gauge \(n=J\).
Using \eqref{eq:BF-curvature-decomposition},
\eqref{eq:sourced-BF-equation} becomes
\begin{equation}
T^a=0,
\qquad
\mathrm d\omega+e^1\wedge e^2
=
2\pi q\,\delta_p.
\label{eq:spin-connection-source}
\end{equation}
Write the Poincar\'e-dual form as
\begin{equation}
\delta_p
=
\delta_p^{(g)}e^1\wedge e^2
=
\delta_p^{(g)}
\sqrt g\,\mathrm d^2x,
\qquad
\int_\Sigma
\sqrt g\,\mathrm d^2x\,
\delta_p^{(g)}(x)
=
1,
\label{eq:scalar-delta-definition}
\end{equation}
where \(\delta_p^{(g)}\) is the covariant scalar delta distribution. From
\begin{equation}
\mathrm d\omega
=
\frac12R\,e^1\wedge e^2,
\qquad
e^1\wedge e^2
=
\sqrt g\,\mathrm d^2x,
\label{eq:curvature-volume-relations-recalled}
\end{equation}
we obtain
\begin{equation}
R+2
=
4\pi q\,\delta_p^{(g)}(x).
\label{eq:conical-curvature-source}
\end{equation}
This is the distributional curvature equation for a hyperbolic surface with
a conical point
\cite{Mertens:2019tcm,Witten:2020wvy,Maxfield:2020ale}. The angular deficit
and opening angle are
\begin{equation}
\delta
=
2\pi q,
\qquad
\theta
=
2\pi(1-q).
\label{eq:defect-angles}
\end{equation}
Near the marked point, the metric may be written as
\begin{equation}
\mathrm ds^2
=
\mathrm dr^2+r^2\mathrm d\varphi^2,
\qquad
\varphi\sim\varphi+2\pi(1-q).
\label{eq:local-conical-metric-q}
\end{equation}
Its singular curvature satisfies
\begin{equation}
\int_{D_\epsilon(p)}
\sqrt g\,\mathrm d^2x\,
R_{\mathrm{sing}}
=
4\pi q.
\label{eq:integrated-singular-curvature}
\end{equation}
The factor of two relative to the angular deficit follows from
\(R=2K\) in two dimensions. Equation
\eqref{eq:conical-curvature-source} agrees with the curvature equation
obtained directly from the metric defect,
\eqref{eq:sourced-curvature-equation}.

The same source fixes the monodromy around the conical point
\cite{Mertens:2019tcm,Iliesiu:2019xuh}. Let \(\gamma_p\) be a small
positively oriented loop around \(p\), and let
\(\phi\sim\phi+2\pi\) be a reference angular coordinate on a punctured
neighborhood of \(p\). In a local gauge in which the singular connection
lies in the Abelian subalgebra generated by \(J\), take
\begin{equation}
A_{\mathrm{sing}}
=
-qJ\,\mathrm d\phi.
\label{eq:singular-local-connection}
\end{equation}
The distributional identity
\begin{equation}
\mathrm d(\mathrm d\phi)
=
2\pi\delta_p
\label{eq:angular-distribution-identity}
\end{equation}
then gives
\begin{equation}
F_{\mathrm{sing}}
=
\mathrm dA_{\mathrm{sing}}
=
-2\pi qJ\,\delta_p,
\label{eq:singular-local-curvature}
\end{equation}
in agreement with \eqref{eq:sourced-BF-equation}. Since
\(A_{\mathrm{sing}}\) lies in a fixed Abelian direction, path ordering is
trivial, and
\begin{equation}
\operatorname{Hol}_{\gamma_p}(A)
\sim
\exp\!\left(
\oint_{\gamma_p}A_{\mathrm{sing}}
\right)
=
\exp(-2\pi qJ).
\label{eq:elliptic-defect-holonomy}
\end{equation}
Here \(\sim\) denotes equality up to conjugation. Reversing the orientation
of \(\gamma_p\) replaces the holonomy by its inverse.

With the normalization \eqref{eq:explicit-sl2-basis}, the compact
generator satisfies
\begin{equation}
J^2
=
-\frac14\mathbf 1.
\label{eq:J-square}
\end{equation}
It therefore generates the rotational one-parameter subgroup
\begin{equation}
\exp(\alpha J)
=
\cos\!\left(\frac{\alpha}{2}\right)\mathbf 1
+
2\sin\!\left(\frac{\alpha}{2}\right)J,
\qquad
\operatorname{Tr}\exp(\alpha J)
=
2\cos\!\left(\frac{\alpha}{2}\right).
\label{eq:J-elliptic-subgroup}
\end{equation}
For \(0<\alpha<2\pi\), the trace has absolute value smaller than two.
Thus \(\exp(\alpha J)\) belongs to an elliptic conjugacy class of
\(\mathrm{SL}(2,\mathbb R)\). In particular,
\begin{equation}
\exp(2\pi J)
=
-\mathbf 1.
\label{eq:J-two-pi-central}
\end{equation}
The deficit-angle and opening-angle representatives are therefore related
by
\begin{equation}
\exp\!\left[
2\pi(1-q)J
\right]
=
-\exp(-2\pi qJ).
\label{eq:deficit-opening-holonomy-relation}
\end{equation}
Their images define the same element of
\(\mathrm{PSL}(2,\mathbb R)\). In \(\mathrm{SL}(2,\mathbb R)\), or in
its universal cover, they correspond to different lifts. We use the lift
\(\exp(-2\pi qJ)\), which is continuously connected to the identity at
\(q=0\).

Thus \(2\pi q\) labels the monodromy by the angular deficit, whereas
\(2\pi(1-q)\) is the geometric opening angle. The opening angle is the
parameter that appears when a cone point is obtained by analytically
continuing a geodesic boundary length
\cite{Witten:2020wvy,Maxfield:2020ale,DoNorbury:2009}.

For defects at distinct fixed points \(p_i\), the source equation
generalizes to
\begin{equation}
F_A
=
-2\pi
\sum_{i=1}^{L}
q_i n_i\,\delta_{p_i}.
\label{eq:multiple-BF-sources}
\end{equation}
In a gravitational gauge with \(n_i=J\), its elliptic component gives
\begin{equation}
R+2
=
4\pi
\sum_{i=1}^{L}
q_i\delta_{p_i}^{(g)}(x),
\label{eq:multiple-conical-sources}
\end{equation}
where
\begin{equation}
\delta_{p_i}
=
\delta_{p_i}^{(g)}
\sqrt g\,\mathrm d^2x.
\label{eq:multiple-scalar-delta}
\end{equation}
For separated marked points, each term produces a conical singularity with
deficit angle \(2\pi q_i\) and opening angle \(2\pi(1-q_i)\). If several
points coincide, their distributional sources add. Whether the fused source
belongs to the compactified gravitational configuration space is a global
question addressed in Section~\ref{sec:fusion}.

We call a defect sharp when
\begin{equation}
q\geq\frac12,
\qquad
\theta\leq\pi,
\label{eq:sharp-defect-definition}
\end{equation}
and blunt when
\begin{equation}
0<q<\frac12,
\qquad
\theta>\pi.
\label{eq:blunt-defect-definition}
\end{equation}
The boundary value \(q=\tfrac12\) is included in the sharp range. Locally it
has the same type of curvature source and elliptic monodromy as the rest of
the family, but globally two identical defects at this value can coincide
only at the cuspidal boundary. The sharp--blunt distinction is therefore
absent from the isolated source equation and becomes relevant only for
integrated multi-defect observables, through the collision strata included
in the compactified position-moduli space
\cite{Turiaci:2020fjj,Eberhardt:2023rzz,Lin:2023trc,
Kruthoff:2024rpx}.

The integrated observable
\begin{equation}
\mathfrak D_q^{\mathrm{BF}}
=
\int_\Sigma
\Omega_n(A)
\exp\!\left(
2\pi iq\langle n,X\rangle
\right)
\label{eq:integrated-defect-source}
\end{equation}
therefore has a direct interpretation. The exponential creates the
localized BF source and fixes its elliptic conjugacy class, while
\(\Omega_n(A)\) integrates that source over its position. For several
defects, the local sources compose, whereas their relative positions remain
global moduli.

\subsection{Defect-generated deformations of JT gravity}
\label{subsec:defect-generated-deformations}

Integrated conical defects generate a class of dilaton-potential
deformations of JT gravity. Such deformations and their matrix-model
description were studied in~\cite{Witten:2020wvy}, and their organization
as a gas of conical defects was developed in
\cite{Maxfield:2020ale,Turiaci:2020fjj}. In our normalization, let
\begin{equation}
I_U[g,\Phi]
=
I_{\mathrm{JT}}[g,\Phi]
-
\frac12
\int_\Sigma
\mathrm d^2x\,\sqrt g\,
U(\Phi).
\label{eq:defect-deformed-JT-action}
\end{equation}
For a finite set of defect species, consider
\begin{equation}
U(\Phi)
=
2\sum_s
\lambda_s e^{-2\pi q_s\Phi},
\qquad
0<q_s<1,
\label{eq:defect-generated-potential}
\end{equation}
where \(q_s\) is the fractional deficit and \(\lambda_s\) is the
corresponding fugacity. The Boltzmann weight then becomes
\begin{align}
e^{-I_U[g,\Phi]}
&=
e^{-I_{\mathrm{JT}}[g,\Phi]}
\exp\!\left[
\sum_s\lambda_s
\int_\Sigma
\mathrm d^2x\,\sqrt g\,
e^{-2\pi q_s\Phi}
\right]
\nonumber\\
&=
e^{-I_{\mathrm{JT}}[g,\Phi]}
\exp\!\left[
\sum_s\lambda_s
\mathfrak D_{q_s}^{\mathrm{metric}}
\right].
\label{eq:defect-potential-generating-functional}
\end{align}
Thus the deformation is equivalently organized as a grand-canonical
expansion in integrated spacetime defects.

The BF observable constructed above provides a gauge-invariant
first-order representative of every insertion in this expansion:
\begin{equation}
\left.
\mathfrak D_{q_s}^{\mathrm{BF}}
\right|_{\Gamma_{\mathrm{grav}}}
=
\mathfrak D_{q_s}^{\mathrm{metric}}.
\label{eq:defect-deformation-BF-dictionary}
\end{equation}
From the BF viewpoint, the local bulk action is left unchanged; the
deformation is introduced through the generating functional of integrated
defect observables, with the path integral restricted to the gravitational
integration cycle.

\section{Exact quantum kernel of the elementary spacetime defect}
\label{sec:quantization}
\label{sec:correlators} % Backward-compatible label used in the original draft.

The preceding section defined the gauge-invariant integrated BF defect
\begin{equation}
\mathfrak D_q^{\mathrm{BF}}
=
\int_\Sigma
\Omega_n(A)\,
\exp\!\left(
2\pi iq\langle n,X\rangle
\right)
\label{eq:section3-integrated-defect}
\end{equation}
and established its reduction on the gravitational integration cycle:
\begin{equation}
\left.
\mathfrak D_q^{\mathrm{BF}}
\right|_{\Gamma_{\mathrm{grav}}}
=
\int_\Sigma
\sqrt g\,\mathrm d^2x\,
e^{-2\pi q\Phi}.
\label{eq:section3-metric-defect}
\end{equation}
The exponential fixes the curvature source and elliptic monodromy, while
\(\Omega_n(A)\) integrates the insertion over its position.

This construction identifies the gravitational meaning of the BF
observable, but it does not by itself rederive the quantum mechanics of the
elliptic sector. The exact quantization of this sector is known from the
twisted-Schwarzian and Virasoro coadjoint-orbit
description~\cite{Mertens:2019tcm}. In this section, we identify the
elliptic sector associated with the BF observable
\eqref{eq:section3-integrated-defect} and express its exact kernel using the
JT spectral and sewing conventions adopted throughout this paper.

We first consider a single defect on the disk. The position integral,
together with the residual quotient that moves the marked point, leaves no
relative-position modulus. The remaining quantum problem is therefore the
fixed elliptic sector. For several defects, only the overall position zero
modes are removed. Their relative positions and collision strata remain
part of the global moduli-space problem discussed in
Section~\ref{sec:fusion}.

\subsection{Quantization strategy}
\label{subsec:quantization-strategy}

The classical BF defect is
\begin{equation}
\mathfrak D_q^{\mathrm{BF}}
=
\int_\Sigma
\Omega_n(A)\,
\exp\!\left(
2\pi iq\langle n,X\rangle
\right).
\label{eq:section3-classical-BF-defect}
\end{equation}
The two factors play different roles. The two-form \(\Omega_n(A)\) provides
the covariant measure for the defect position, while
\begin{equation}
\mathcal O_q(p)
=
\exp\!\left[
2\pi iq\langle n(p),X(p)\rangle
\right]
\label{eq:section3-local-defect-operator}
\end{equation}
fixes the local elliptic source. On the gravitational integration cycle,
\begin{equation}
\left.
\mathcal O_q(p)
\right|_{\Gamma_{\mathrm{grav}}}
=
e^{-2\pi q\Phi(p)}.
\label{eq:section3-local-defect-grav}
\end{equation}

For one defect on the disk, the position modes are removed by the residual
transformations that move the marked point. After this quotient, the defect
may be placed at a reference point \(p_0\). The position form
\(\Omega_n(A)\) contributes to the normalization of this reduction, while
the remaining quantum problem is the quantization of
\(\mathcal O_q(p_0)\). Section~\ref{subsec:fixing-defect-position} describes
the position quotient explicitly.

The BF source equation fixes the opening angle
\begin{equation}
\theta_q
=
2\pi(1-q),
\qquad
a_q=1-q.
\label{eq:section3-opening-parameter}
\end{equation}
The reduced disk problem is therefore the elliptic sector labelled by
\(a_q\). Its exact boundary quantization is known from the twisted
Schwarzian and Virasoro coadjoint-orbit description
\cite{Mertens:2019tcm}. We will identify the corresponding reduced matrix
element in the JT spectral convention and use it to compute the exact
one-defect disk amplitude.

\subsection{JT boundary spectrum and the reduced defect operator}
\label{subsec:boundary-quantization}

The asymptotic JT boundary condition turns the otherwise topological BF
theory into the Schwarzian boundary quantum mechanics
\cite{Blommaert:2018oue,Iliesiu:2019xuh}. Away from defect insertions,
\begin{equation}
F_A=0,
\qquad
D_AX=0,
\label{eq:section3-bulk-constraints}
\end{equation}
and the physical states are organized into
\(\mathrm{SL}(2,\mathbb R)\) representation channels.

The continuous sector relevant to the JT disk is parametrized by
\begin{equation}
k\in\mathbb R_{\geq0},
\qquad
C_2(k)=k^2+\frac14.
\label{eq:principal-series-data}
\end{equation}
We choose
\begin{equation}
H=C_2-\frac14,
\qquad
E(k)=k^2.
\label{eq:section3-boundary-energy}
\end{equation}
With the topological factor \(e^{S_0}\) removed, the JT disk amplitude is
\begin{equation}
Z_{\mathrm{JT}}(\beta)
=
\frac{1}{4\sqrt{\pi}\,\beta^{3/2}}
\exp\!\left(
\frac{\pi^2}{\beta}
\right)
=
\int_0^\infty
\mathrm dE\,
\rho_{\mathrm{JT}}(E)e^{-\beta E},
\label{eq:JT-disk-normalization}
\end{equation}
where
\begin{equation}
\rho_{\mathrm{JT}}(E)
=
\frac{\sinh(2\pi\sqrt E)}{4\pi^2}.
\label{eq:JT-spectral-density}
\end{equation}
Equivalently,
\begin{equation}
Z_{\mathrm{JT}}(\beta)
=
\int_0^\infty
\mathrm dk\,
w_{\mathrm{JT}}(k)e^{-\beta k^2},
\qquad
w_{\mathrm{JT}}(k)
=
\frac{k\sinh(2\pi k)}{2\pi^2}.
\label{eq:JT-disk-k-representation}
\end{equation}

We now insert the fixed-position operator
\begin{equation}
\mathcal O_q(p_0)
=
\exp\!\left[
2\pi iq\langle n(p_0),X(p_0)\rangle
\right].
\label{eq:section3-fixed-position-operator}
\end{equation}
In canonical BF quantization, the components of \(X\) act as
\(\mathfrak{sl}(2,\mathbb R)\) generators. Since the quadratic Casimir is
central,
\begin{equation}
[C_2,X^a]=0,
\end{equation}
and therefore
\begin{equation}
[H,\widehat{\mathcal O}_q]=0.
\label{eq:defect-Hamiltonian-commutator}
\end{equation}
The defect preserves the Casimir label \(k\), so its matrix elements have
the form
\begin{equation}
\langle k,\alpha|
\widehat{\mathcal O}_q
|k',\alpha'\rangle
=
\delta(k-k')\,
\mathcal O^{(q)}_{\alpha\alpha'}(k),
\label{eq:fixed-defect-block-diagonal}
\end{equation}
up to the normalization of the continuous states.

The commutator with \(H\) does not imply that
\(\mathcal O^{(q)}_{\alpha\alpha'}(k)\) is proportional to the identity
within the full principal-series representation. In the gravitational disk
amplitude, however, the JT boundary condition and the quotient by the
defect position and elliptic stabilizer contract the remaining internal
indices. The resulting scalar matrix element will be denoted by
\begin{equation}
K_q(k).
\label{eq:defect-kernel-definition}
\end{equation}

This scalar appears naturally from BF sewing. Cutting the disk along a
closed curve \(\gamma\) that separates the defect from the asymptotic
boundary produces an interior defect disk and an exterior annulus. Sewing
the two pieces in a common representation channel gives
\begin{equation}
Z_{1\text{-}\mathrm{def}}(\beta;q)
=
\int_0^\infty
\mathrm dk\,
w_{\mathrm{JT}}(k)\,
e^{-\beta k^2}\,
K_q(k).
\label{eq:disk-defect-spectral-k}
\end{equation}
Here \(w_{\mathrm{JT}}(k)\,\mathrm dk\) is the BF/JT sewing measure,
\(e^{-\beta k^2}\) is the Schwarzian evolution along the asymptotic
boundary, and \(K_q(k)\) is the reduced elliptic matrix element.

Equation~\eqref{eq:disk-defect-spectral-k} determines the form of the
one-defect amplitude once \(K_q(k)\) is known. The position quotient is
analyzed in Section~\ref{subsec:fixing-defect-position}, and the exact
twisted-Schwarzian quantization of the resulting elliptic sector is used in
Section~\ref{subsec:exact-elementary-defect} to determine \(K_q(k)\).

The limit \(q\to0\) should not be confused with removing the insertion.
Although the curvature source vanishes, the integrated BF operator still
contains the position integral of a smooth marked point. Its normalization
therefore differs from that of the ordinary JT cap, and in particular one
should not impose
\begin{equation}
K_{q=0}(k)=1.
\end{equation}

\subsection{Fixing the defect position}
\label{subsec:fixing-defect-position}

We now explain how the integrated observable
\eqref{eq:section3-integrated-defect} is related, for one defect on the disk,
to the fixed elliptic sector entering
\eqref{eq:disk-defect-spectral-k}.

For a chosen hyperbolic disk, the residual
\(\mathrm{PSL}(2,\mathbb{R})\) action is transitive on the bulk. After
choosing a reference point \(p_0\), its orbit is
\begin{equation}
\mathbb{H}^2
\simeq
\frac{\mathrm{PSL}(2,\mathbb{R})}{U(1)},
\label{eq:hyperbolic-disk-coset}
\end{equation}
where \(U(1)\) is the subgroup that fixes \(p_0\). Consequently,
the two coordinates of a single marked point do not define relative moduli:
they may be fixed by two gauge conditions, with the corresponding
Faddeev--Popov determinant and stabilizer normalization included. In the
boundary description, the same reorganization replaces
\(\mathrm{Diff}(S^1)/\mathrm{PSL}(2,\mathbb{R})\), together with the
position integral, by the elliptically twisted orbit
\(\mathrm{Diff}(S^1)/U(1)\)
\cite{Mertens:2019tcm,Lin:2023trc}.

The relation between the position measure and the BF gauge orbit can be
seen locally. Let \(y^\mu\) denote the coordinates of the marked point. On
the gravitational integration cycle, an infinitesimal displacement
\(\delta y^\mu\) is related on shell to the translational part of a BF gauge
transformation by
\begin{equation}
\delta\xi^a
=
e^a{}_{\mu}(p)\,\delta y^\mu.
\label{eq:translation-diffeomorphism-map}
\end{equation}
The Jacobian of this change of variables is
\begin{equation}
\det\!\left(
\frac{\partial\xi^a}{\partial y^\mu}
\right)
=
\det(e^a{}_{\mu})
=
\sqrt{g},
\label{eq:position-Jacobian}
\end{equation}
where the orientation condition defining
\(\Gamma_{\mathrm{grav}}\) has been used. Hence
\begin{equation}
\mathrm{d}\xi^1\,\mathrm{d}\xi^2
=
\sqrt{g}\,\mathrm{d}^2y.
\label{eq:position-measure-matching}
\end{equation}
The invariant measure along the two directions that move the defect
therefore reduces to the gravitational area measure of the marked point.

This is precisely the measure supplied by the BF two-form:
\begin{equation}
\left.
\Omega_n(A)
\right|_{\Gamma_{\mathrm{grav}}}
=
e^1\wedge e^2
=
\sqrt{g}\,\mathrm{d}^2y.
\label{eq:Omega-position-measure}
\end{equation}
Thus \(\Omega_n(A)\) has a concrete role in the reduction: it is the
covariant measure along the two-dimensional position orbit. It converts the
fixed local exponential into an integrated spacetime observable.

Formally, the relevant part of the gauge-fixed path integral contains the
combination
\begin{equation}
\frac{1}{\operatorname{Vol}(G)}
\int_{G/H}
\mathrm{d}\mu_{G/H}(p)\,
\exp\!\left[
2\pi iq\langle n(p),X(p)\rangle
\right],
\qquad
G=\mathrm{PSL}(2,\mathbb{R}),
\qquad
H=U(1).
\label{eq:position-orbit-quotient}
\end{equation}
The integral over \(G/H\) is paired with the part of the residual gauge
quotient that moves the point. After fixing \(p=p_0\), only the compact
stabilizer and the normalization of the position zero modes remain. Since
\(G\) and \(G/H\) are noncompact, this statement is understood at the level
of the gauge-fixed measure; it is not an equality of separately defined
group volumes.

The resulting fixed-position representation may be written as
\begin{equation}
\mathfrak{D}^{\mathrm{BF}}_q
\quad\longrightarrow\quad
\mathcal{N}_{\mathrm{pos}}(q)
\exp\!\left[
2\pi iq
\langle n(p_0),X(p_0)\rangle
\right].
\label{eq:fixed-position-defect}
\end{equation}
Here \(\mathcal{N}_{\mathrm{pos}}(q)\) summarizes the normalization of the
two position modes, the Faddeev--Popov determinant, the
\(U(1)\) stabilizer, and the BF sewing states. Equation
\eqref{eq:position-measure-matching} fixes the local Jacobian entering this
factor, but it does not by itself determine the complete finite
normalization. In particular, the norms of the position modes can depend on
the conical background, as is explicit in the gauge-fixed metric treatment
\cite{Lin:2023trc}. We therefore do not regard
\(\mathcal{N}_{\mathrm{pos}}(q)\) as an independent coupling; it is absorbed
into the normalized elliptic kernel \(K_q(k)\), whose convention is fixed by
the JT disk and sewing amplitudes.

On \(\Gamma_{\mathrm{grav}}\), choose the local gravitational gauge \(n=J\).
The contour condition
\begin{equation}
\langle J,X\rangle=i\Phi
\end{equation}
then gives
\begin{equation}
\exp\!\left[
2\pi iq\langle J,X(p_0)\rangle
\right]
=
e^{-2\pi q\Phi(p_0)}.
\label{eq:fixed-position-metric-insertion}
\end{equation}
The reduced insertion therefore creates the same fixed conical source and
elliptic monodromy as the sector quantized in
\cite{Mertens:2019tcm}.

The identification is made only after the position quotient. Before gauge
fixing, \(\mathfrak D_q^{\mathrm{BF}}\) is an integrated gauge-invariant
spacetime observable, and \(\Omega_n(A)\) is part of its definition. After
the position orbit has been removed, the remaining data are the elliptic
parameter and the normalization of the corresponding fixed sector. It is
this reduced problem to which the exact twisted-Schwarzian kernel applies.

This simplification is special to one marked point on the disk. With several
defects, the residual symmetry removes only the overall position zero modes.
Relative positions remain genuine moduli, and their measures and collision
strata are not contained in the elementary kernel \(K_q(k)\). They form the
global localization problem studied in Section~\ref{sec:fusion}.

\subsection{Elliptic-orbit kernel and one-defect amplitude}
\label{subsec:exact-elementary-defect}

The defect has opening angle
\begin{equation}
\theta_q
=
2\pi(1-q),
\qquad
a_q
\equiv
\frac{\theta_q}{2\pi}
=
1-q.
\label{eq:opening-angle-quantum}
\end{equation}
After the position quotient described in
Section~\ref{subsec:fixing-defect-position}, the remaining degree of freedom
is therefore the elliptic sector \(U(1)_{a_q}\).

The exact quantization of this sector was obtained from the twisted
Schwarzian and the corresponding Virasoro coadjoint orbit in
\cite{Mertens:2019tcm}. Its dependence on the principal-series parameter is
described by
\begin{equation}
D_{U(1)_a}(k)
=
\frac{\cosh(2\pi ak)}
{k\sinh(2\pi k)}.
\label{eq:MT-elliptic-defect-factor}
\end{equation}
To use this result in the JT spectral convention of
Section~\ref{subsec:boundary-quantization}, its overall normalization must be
converted to the disk and trumpet sewing convention adopted here. This gives
\begin{equation}
K_q(k)
=
2\pi D_{U(1)_{1-q}}(k)
=
\frac{
2\pi\cosh\!\left(2\pi(1-q)k\right)
}{
k\sinh(2\pi k)
}.
\label{eq:exact-defect-kernel}
\end{equation}
The factor \(2\pi\) is fixed by the standard JT trumpet normalization below.
It fixes the joint convention for the normalized defect operator and its
fugacity; rescaling one requires the inverse rescaling of the other.

Combining \eqref{eq:exact-defect-kernel} with the JT spectral weight
\eqref{eq:JT-disk-k-representation} gives
\begin{equation}
w_{\mathrm{JT}}(k)K_q(k)
=
\frac{1}{\pi}
\cosh\!\left(2\pi(1-q)k\right).
\label{eq:measure-kernel-cancellation}
\end{equation}
The sewing formula \eqref{eq:disk-defect-spectral-k} consequently becomes
\begin{equation}
Z_{1\text{-}\mathrm{def}}(\beta;q)
=
\frac{1}{\pi}
\int_0^\infty
\mathrm{d}k\,
\cosh\!\left(2\pi(1-q)k\right)
e^{-\beta k^2}.
\label{eq:defect-amplitude-k-integral}
\end{equation}

This normalization can be stated geometrically in terms of the exact JT
trumpet. For a geodesic boundary of length \(b\),
\begin{equation}
Z_{\mathrm{tr}}(\beta,b)
=
\frac{1}{\pi}
\int_0^\infty
\mathrm{d}k\,
\cos(bk)e^{-\beta k^2}
=
\frac{1}{2\sqrt{\pi\beta}}
\exp\!\left(
-\frac{b^2}{4\beta}
\right)
\label{eq:exact-trumpet-amplitude}
\end{equation}
in the convention used here
\cite{Saad:2019lba}. A cone point of opening angle \(\theta_q\) is obtained
by the elliptic continuation
\begin{equation}
b=i\theta_q=2\pi i(1-q).
\label{eq:elliptic-continuation}
\end{equation}
The continued trumpet integrand is precisely the integrand in
\eqref{eq:defect-amplitude-k-integral}. Notice that the continuation involves
the opening angle, not the deficit angle: the cusp limit is
\(\theta_q\to0\), or \(q\to1\), while \(q\to0\) is the smooth
marked-point endpoint. The relation between imaginary boundary length and
conical JT amplitudes is also discussed in
\cite{Witten:2020wvy,Maxfield:2020ale,Eberhardt:2023rzz,Lin:2023trc}.

In the energy variable \(E=k^2\), the corresponding spectral density is
\begin{equation}
\rho_q(E)
=
\frac{
\cosh\!\left(
2\pi(1-q)\sqrt{E}
\right)
}{
2\pi\sqrt{E}
},
\qquad
E>0,
\label{eq:exact-defect-spectral-density}
\end{equation}
and the equality of the two spectral descriptions is
\begin{equation}
w_{\mathrm{JT}}(k)K_q(k)\,\mathrm{d}k
=
\rho_q(E)\,\mathrm{d}E.
\label{eq:kernel-density-matching}
\end{equation}
Thus the twisted-Schwarzian elliptic kernel, after conversion to the JT
sewing convention, is the same kernel obtained by elliptically continuing
the exact trumpet.

Evaluating the Gaussian integral gives
\begin{equation}
Z_{1\text{-}\mathrm{def}}(\beta;q)
=
\frac{1}{2\sqrt{\pi\beta}}
\exp\!\left[
\frac{\pi^2(1-q)^2}{\beta}
\right].
\label{eq:exact-one-defect-amplitude}
\end{equation}
Relative to the undeformed JT disk amplitude
\eqref{eq:JT-disk-normalization}, this may be written as
\begin{equation}
Z_{1\text{-}\mathrm{def}}(\beta;q)
=
2\beta\,Z_{\mathrm{JT}}(\beta)
\exp\!\left[
-\frac{\pi^2q(2-q)}{\beta}
\right].
\label{eq:one-defect-relative-amplitude}
\end{equation}
This is the elementary one-defect contribution used in the later
multi-defect expansion. It agrees with the corresponding first-order term in
deformed JT gravity
\cite{Witten:2020wvy,Maxfield:2020ale,Turiaci:2020fjj}.

\subsection{Classical comparison and scope}
\label{subsec:section3-comparison}

We finally compare the normalized one-defect result with the direct evaluation of the metric insertion on the undeformed disk saddle, following the analysis of~\cite{Witten:2020wvy}, in particular Sections~2 and~5. In the
conventions of Section~\ref{subsec:jt-and-defects}, the latter is
\begin{equation}
\mathrm{d}s^2
=
\mathrm{d}\rho^2
+
\sinh^2\rho\,\mathrm{d}\psi^2,
\qquad
\psi\sim\psi+2\pi,
\qquad
\Phi(\rho)
=
\frac{\pi}{\beta}\cosh\rho.
\label{eq:hyperbolic-disk-saddle-comparison}
\end{equation}
The integrated metric insertion then gives
\begin{align}
\mathfrak{D}^{\mathrm{cl}}_q
&=
\int_0^{2\pi}\mathrm{d}\psi
\int_0^\infty\mathrm{d}\rho\,
\sinh\rho\,
\exp\!\left[
-\frac{2\pi^2q}{\beta}\cosh\rho
\right]
\nonumber\\
&=
\frac{\beta}{\pi q}
\exp\!\left(
-\frac{2\pi^2q}{\beta}
\right).
\label{eq:classical-defect-comparison}
\end{align}
For \(q>0\), the dilaton exponential suppresses the asymptotic region, so
this position integral is finite. It is nevertheless only the evaluation of
the bare insertion on the undeformed saddle; it does not include the change
of saddle or the quantum normalization of the integrated operator.

The corresponding normalized quantum expectation value follows from
\eqref{eq:one-defect-relative-amplitude}:
\begin{equation}
\frac{
Z_{1\text{-}\mathrm{def}}(\beta;q)
}{
Z_{\mathrm{JT}}(\beta)
}
=
2\beta
\exp\!\left[
-\frac{\pi^2q(2-q)}{\beta}
\right].
\label{eq:exact-defect-comparison}
\end{equation}
Compared with \eqref{eq:classical-defect-comparison}, the exponent changes
according to
\begin{equation}
-\frac{2\pi^2q}{\beta}
\quad\longrightarrow\quad
-\frac{2\pi^2q}{\beta}
+
\frac{\pi^2q^2}{\beta},
\label{eq:exponent-quantum-shift}
\end{equation}
while the prefactor, in the fugacity convention used here, changes as
\begin{equation}
\frac{\beta}{\pi q}
\quad\longrightarrow\quad
2\beta.
\label{eq:normalization-quantum-shift}
\end{equation}
The \(q^2\) term is contained in the exact elliptic kernel and accounts for
the replacement of the undeformed disk saddle by the conical sector. The
prefactor is fixed by the normalized position quotient and the JT sewing
convention. These effects modify the quantum amplitude without changing the
classical source equation.

The limit \(q\to0\) makes the distinction between the bare and normalized
operators particularly clear. At the level of the metric insertion,
\begin{equation}
e^{-2\pi q\Phi}
\longrightarrow
1,
\end{equation}
and hence
\begin{equation}
\mathfrak{D}^{\mathrm{metric}}_q
\longrightarrow
\int_\Sigma
\sqrt{g}\,\mathrm{d}^2x.
\label{eq:q-zero-bare-area}
\end{equation}
The asymptotic region makes this bare area divergent. Indeed,
\eqref{eq:classical-defect-comparison} has the expansion
\begin{equation}
\mathfrak{D}^{\mathrm{cl}}_q
=
\frac{\beta}{\pi q}
-
2\pi
+
\mathcal{O}(q),
\qquad
q\to0.
\label{eq:q-zero-classical-expansion}
\end{equation}
The pole comes from the large-area asymptotic region. More importantly, the
limit \(q\to0\) is not uniformly captured by the classical saddle-point
approximation. As the dilaton suppression of the asymptotic region
disappears, the position modes become increasingly important. Their measure
must therefore be treated together with the fluctuation determinant and the
elliptic stabilizer. A classical evaluation of the bare insertion is
insufficient in this limit; at minimum, one must also include the one-loop
and collective-coordinate contributions. We therefore use the exact elliptic kernel instead of expanding around the degenerating \(q\to0\)
saddle.

The additional \(q^2\) term in the exact exponent vanishes in this limit,
but the discrepancy in the prefactor remains. A subtraction of the
divergent area from the classical expression would leave the finite term
\(-2\pi\), whereas the normalized quantum insertion approaches \(2\beta\).
Thus, the smooth marked-point operator cannot be defined by a geometric
subtraction of the classical area alone. Its finite normalization is fixed
by continuing the exact elliptic kernel in the JT sewing convention. In the
spectral convention adopted above, we define the smooth endpoint by
continuing the normalized elliptic family:
\begin{equation}
\widehat{\mathfrak D}^{\,\mathrm{ren}}_0
\equiv
\lim_{q\to0^+}
\widehat{\mathfrak D}_q.
\label{eq:renormalized-q-zero-defect}
\end{equation}
Equation~\eqref{eq:exact-defect-comparison} then gives
\begin{equation}
\lim_{q\to0^+}
\frac{
Z_{1\text{-}\mathrm{def}}(\beta;q)
}{
Z_{\mathrm{JT}}(\beta)
}
=
2\beta.
\label{eq:q-zero-exact-amplitude}
\end{equation}
The limit is finite, but it is not the amplitude obtained by removing the
operator. At \(q=0\) the curvature source vanishes, while the integration
over the marked point remains:
\begin{equation}
\mathfrak D^{\mathrm{BF}}_0
=
\int_\Sigma\Omega_n(A).
\label{eq:q-zero-marked-point}
\end{equation}
Thus
\begin{equation}
\widehat{\mathfrak D}^{\,\mathrm{ren}}_0
\neq
\mathbf{1}.
\label{eq:q-zero-not-identity}
\end{equation}
It represents a renormalized smooth marking rather than the unmarked JT
disk.

The result \eqref{eq:one-defect-relative-amplitude} agrees with the
one-defect term obtained in deformed JT gravity and its matrix-model
description
\cite{Witten:2020wvy,Maxfield:2020ale,Turiaci:2020fjj}. This comparison
checks both the angular dependence and the joint normalization of the
defect operator and its fugacity.

The ingredients entering this result should be distinguished. The BF
observable constructed in Section~\ref{subsec:BF-defect} determines the
position measure, the localized curvature source, and the elliptic
conjugacy class of the holonomy around the moving defect. On the disk, the
quotient by transformations that move the defect removes its position
modulus and leaves the \(U(1)_{1-q}\) elliptic sector. The exact
\(k\)-dependence of this sector is supplied by the twisted-Schwarzian
quantization of~\cite{Mertens:2019tcm} and is translated here into our
JT spectral convention. We do not derive it independently by evaluating a
BF orbital integral.

All statements in this section refer to the gravitational integration cycle
\(\Gamma_{\mathrm{grav}}\). Moreover, the one-defect disk has no relative
position modulus. For several defects, only the overall position modes are
removed, while relative positions and their collision limits remain. The
elementary kernel found here is common to sharp and blunt defects; their
different multi-defect amplitudes arise from the compactification of those
relative-position moduli, which is the subject of the next section.

\section{Gauge-theoretic origin of defect moduli-space geometry}
\label{sec:fusion}

In the previous section, we identified the elementary BF defect with the
corresponding exactly quantized elliptic sector and determined its kernel in
the JT spectral and sewing conventions used in this paper. We now turn to
the global geometry that arises when several moving defects are present.

The key observation is that the gauge-invariant BF defect operators are
integrated over their insertion points. After localization on the
gravitational integration cycle, these insertion points become marked
conical points on a hyperbolic surface. The reduced BF measure on the gravitational component is the
Weil--Petersson measure; see
Appendix~\ref{app:moment_map_reduction} for its derivation and normalization. The localized path integral
therefore naturally leads to a moduli problem for hyperbolic surfaces with
moving cone points. In the conventional geometric formulation, the
corresponding moduli spaces and their Weil--Petersson geometry provide the
basis for evaluating JT amplitudes with conical
defects~\cite{Witten:2020wvy,Maxfield:2020ale,Eberhardt:2023rzz,
Lin:2023trc}. Our aim is to identify the gauge-theoretic origin of the
marked-point data and the local collision criterion that enters the
compactification of these moduli spaces.

Two complementary structures will be important. The first is associated
with an individual moving defect. A marked point \(p_i\) on a family of
curves carries a cotangent-line bundle
\begin{equation}
\mathbb{L}_i
\longrightarrow
\overline{\mathcal{M}}_{g,L},
\qquad
\left.
\mathbb{L}_i
\right|_{[\Sigma,p_1,\ldots,p_L]}
=
T_{p_i}^{*}\Sigma,
\label{eq:cotangent-line-bundle-preview}
\end{equation}
whose first Chern class is the corresponding \(\psi\)-class,
\begin{equation}
\psi_i
=
c_1(\mathbb{L}_i).
\label{eq:psi-class-preview}
\end{equation}
These classes are standard ingredients in the intersection theory of moduli
spaces of pointed curves
\cite{Witten:1990hr,Kontsevich:1992ti}. We will argue that, in the localized
BF description, the cotangent-line bundle is induced by the residual
rotational symmetry at the defect. The associated \(U(1)\) bundle thereby
provides a gauge-theoretic interpretation of the class \(\psi_i\).

The second structure is global and concerns collisions among several moving
defects. For a given set of Hassett weights, marked points are allowed to
coincide when their total weight does not exceed one, subject also to the
global weighted-stability condition on each component~\cite{Hassett:2003}. The moduli spaces of hyperbolic cone surfaces exhibit the same chamber
structure, and their natural compactifications are Hassett spaces. The different weight chambers are related by reduction morphisms. When a
collection of markings becomes admissible to collide, the corresponding
morphism contracts the resolved rational tail to a point carrying the
coincident weighted markings.\cite{Anagnostou:2022cmq,Anagnostou:2023jiy}. In JT gravity, the
corresponding collision strata, contact terms, and modifications of the
Weil--Petersson volumes enter the treatment of defects with general opening angles~\cite{Turiaci:2020fjj,Eberhardt:2023rzz,Lin:2023trc,
Kruthoff:2024rpx}.

In the BF description, the weights of the localized sources add when their
insertion points coincide. This local source-composition rule does not by
itself determine whether the resulting configuration belongs to the
gravitational localization space. The additional restriction is imposed by
\(\Gamma_{\mathrm{grav}}\): the fused source remains in the compactified
hyperbolic cone sector only when its effective opening angle is
nonnegative. A positive opening angle gives an ordinary cone point, while a
vanishing opening angle gives the cuspidal limit. We will show that this
condition reproduces the Hassett coincidence rule. The identification of
the full compactification further requires the global weighted-stability
condition and the known geometry of the moduli spaces of cone surfaces.

The remainder of this section develops this correspondence. We first
explain how moving BF defects give rise to marked-point moduli and how their
residual rotational symmetry determines the cotangent-line bundles. We then
derive the composition rule for localized BF sources and determine which
collision channels are admitted by the gravitational integration cycle.
Finally, we match the resulting local collision criterion with the weighted
diagonal strata of the Hassett compactification and discuss the associated
conical Weil--Petersson geometry.

\subsection{BF localization and moving defects}
\label{subsec:BF-localization-moving-defects}

Section~\ref{sec:quantization} related the reduced one-defect problem to the
exactly quantized elliptic sector and determined its kernel in the JT
spectral and sewing convention. The gravitational observable, however, is
not a fixed-position insertion. It is the integrated BF operator
\begin{equation}
\mathfrak D_q^{\mathrm{BF}}
=
\int_\Sigma
\Omega_n(A)\,
\mathcal O_q,
\qquad
\mathcal O_q
=
\exp\!\left(
2\pi iq\langle n,X\rangle
\right).
\label{eq:moving-defect}
\end{equation}
The exponential fixes the local elliptic source, while
\(\Omega_n(A)\) supplies the covariant measure over its position.

On the gravitational integration cycle, infinitesimal diffeomorphisms are
related on shell to BF gauge transformations
\cite{Isler:1989hq,Chamseddine:1989yz,Grumiller:2002nm}. A coordinate
displacement of a marked point must therefore be considered together with
the corresponding transformation of the BF fields. The physical
position data are what remain after the combined configuration of fields
and insertion points has been divided by these transformations.

The disk with one defect is a special case. Its residual
\(\mathrm{PSL}(2,\mathbb R)\) symmetry acts transitively on the bulk, and
the stabilizer of a reference point is \({U}(1)\):
\begin{equation}
\mathbb H^2
\simeq
\frac{\mathrm{PSL}(2,\mathbb R)}
{{U}(1)}.
\label{eq:section4-hyperbolic-disk-coset}
\end{equation}
The integration over the defect position can therefore be combined with
the two noncompact directions removed by the residual
\(\mathrm{PSL}(2,\mathbb R)\) quotient. At the level of the gauge-fixed
path-integral measure, this reorganizes the quotient as
\begin{equation}
\mathbb H^2
\times
\frac{\mathrm{Diff}(S^1)}
{\mathrm{PSL}(2,\mathbb R)}
\;\longrightarrow\;
\frac{\mathrm{Diff}(S^1)}
{{U}(1)}.
\label{eq:single-defect-position-reorganization}
\end{equation}
This is the position-orbit reduction described in
Section~\ref{subsec:fixing-defect-position}. The relation is understood at
the level of the gauge-fixed measure, including the zero-mode and
stabilizer normalizations, rather than as an identity between ordinary
finite-dimensional volumes. The corresponding metric calculation and its
angle-dependent position measure were analyzed in~\cite{Lin:2023trc}.

For several defects, the residual \(\mathrm{PSL}(2,\mathbb R)\) acts
diagonally on all insertion points. It removes the overall position modes
of the configuration, but the relative positions remain. On a fixed
hyperbolic disk, these degrees of freedom are locally modeled by
\begin{equation}
\frac{(\mathbb H^2)^L}
{\mathrm{PSL}(2,\mathbb R)},
\label{eq:relative-position-quotient}
\end{equation}
with configurations having nontrivial stabilizers treated separately.
This quotient is only a local model for the relative-position degrees of
freedom. In the gravitational path integral, the defect positions and the
hyperbolic geometry are integrated simultaneously.

After imposing the sourced BF equations on
\(\Gamma_{\mathrm{grav}}\), a separated configuration is described by a
hyperbolic surface carrying labelled conical points,
\begin{equation}
(\Sigma,p_1,\ldots,p_L),
\label{eq:localized-marked-surface}
\end{equation}
with opening angles
\begin{equation}
\theta_i
=
2\pi(1-q_i).
\label{eq:localized-cone-angles}
\end{equation}
The finite-dimensional localization locus is therefore the corresponding
moduli space of hyperbolic cone surfaces. This is the geometric setting used
in the Weil--Petersson description of JT gravity with conical defects
\cite{Witten:2020wvy,Maxfield:2020ale,Eberhardt:2023rzz,Lin:2023trc}.

From the BF viewpoint, the marked points are not added after localization as independent external data. They descend from the position integrations
already present in the operators
\(\mathfrak D_{q_i}^{\mathrm{BF}}\). The gauge quotient removes the overall motion of the configuration, while the relative positions of the defects remain as genuine moduli.

This description initially applies on the open locus where the conical
points are distinct. The integrated correlator also requires the
degeneration limits in which several marked points approach one another.
The position quotient alone does not determine which of these limits belong
to the compactified gravitational localization space. That question depends
on the composition of the localized BF sources and on whether the resulting
geometry remains on \(\Gamma_{\mathrm{grav}}\). We now turn to these
conditions.

\subsection{Composition of localized BF sources}
\label{subsec:source-composition}

For \(L\) defects at distinct points \(p_i\), integration over the BF scalar
imposes
\begin{equation}
F_A
=
-2\pi
\sum_{i=1}^{L}
q_i n_i\,\delta_{p_i},
\label{eq:multi-source}
\end{equation}
where \(q_i\) is the fractional deficit, \(n_i\) is the local elliptic source
direction, and \(\delta_{p_i}\) is the Poincar\'e-dual two-form of \(p_i\).
Away from the marked points, the connection is flat. On the gravitational
integration cycle, \eqref{eq:multi-source} is the first-order form of the
distributional curvature equation for a hyperbolic surface with conical
points
\cite{Witten:2020wvy,Maxfield:2020ale,Eberhardt:2023rzz}.

Consider a subset
\begin{equation}
I\subset\{1,\ldots,L\}
\end{equation}
whose insertion points approach a common point \(p_I\). To compare the
adjoint-valued sources, choose a local trivialization in a neighborhood of
the collision. Equivalently, parallel transport their directions to
\(p_I\). At the level of distributions,
\begin{equation}
\sum_{i\in I}
q_i n_i\,\delta_{p_i}
\longrightarrow
\left(
\sum_{i\in I}
q_i n_i(p_I)
\right)\delta_{p_I}.
\label{eq:cluster-source}
\end{equation}

The gravitational defects considered here belong to the same elliptic
reduction. In a neighborhood of \(p_I\), one may therefore choose a gauge in
which
\begin{equation}
n_i=J,
\qquad
i\in I.
\label{eq:aligned-source-directions}
\end{equation}
The cluster source then becomes
\begin{equation}
\sum_{i\in I}
q_i n_i\,\delta_{p_i}
\longrightarrow
q_IJ\,\delta_{p_I},
\qquad
q_I
\equiv
\sum_{i\in I}q_i.
\label{eq:aligned-cluster-source}
\end{equation}
The sourced BF equation in the collision limit is consequently
\begin{equation}
F_A
=
-2\pi q_IJ\,\delta_{p_I}
-
2\pi
\sum_{j\notin I}
q_jn_j\,\delta_{p_j}.
\label{eq:qI}
\end{equation}
Thus the fractional deficits of aligned gravitational sources add:
\begin{equation}
q_I
=
\sum_{i\in I}q_i.
\label{eq:additive-source}
\end{equation}
The same composition law is visible directly at the level of the local
holonomy. In the common gravitational reduction, each defect has elliptic
holonomy
\begin{equation}
U_i
\sim
\exp(-2\pi q_i J).
\end{equation}
When the defects approach one another, the holonomy around a small loop
enclosing the entire cluster is the product of the individual holonomies.
Since all sources are aligned along the same generator \(J\), these
holonomies commute, and hence
\begin{equation}
U_I
\sim
\prod_{i\in I}
\exp(-2\pi q_iJ)
=
\exp(-2\pi q_IJ),
\qquad
q_I=\sum_{i\in I}q_i.
\label{eq:aligned-holonomy-composition}
\end{equation}
This is the monodromy associated with the fused source in
\eqref{eq:qI}. The statement applies to the local collision limit; for
punctures at finite separation, the holonomies must be transported to a
common base point before they are composed.

The alignment condition is essential. For arbitrary source directions, the
coincident distribution would be characterized by the Lie-algebra element
\begin{equation}
Q_I
=
\sum_{i\in I}
q_i n_i(p_I),
\label{eq:generic-cluster-charge}
\end{equation}
whose conjugacy class depends on
\begin{equation}
\langle Q_I,Q_I\rangle
=
\sum_{i,j\in I}
q_iq_j
\langle n_i(p_I),n_j(p_I)\rangle.
\label{eq:cluster-Casimir}
\end{equation}
It is therefore not determined by the scalar sum
\(\sum_iq_i\) and need not describe an elementary conical defect of the
gravitational family. Equation~\eqref{eq:additive-source} is specifically
the composition law for sources belonging to the common gravitational
elliptic reduction.

This is a local statement about the sourced BF equation. It identifies the
parameter of a candidate fused defect, but it does not yet imply that the
coincident configuration belongs to the compactified gravitational
localization space. Nor does it determine the coefficient with which the
corresponding collision locus contributes to an integrated correlator. The
first question is governed by the gravitational admissibility condition
derived in the next subsection; the second requires the compactified
Weil--Petersson measure and its contact terms
\cite{Turiaci:2020fjj,Eberhardt:2023rzz,Lin:2023trc,
Kruthoff:2024rpx}.

\subsection{Effective conical defects and admissible collisions}
\label{subsec:admissible}

The \(i\)-th defect has opening angle
\begin{equation}
\theta_i
=
2\pi(1-q_i).
\label{eq:individual-opening-angle}
\end{equation}
For an aligned cluster \(I\), the source-composition law
\eqref{eq:additive-source} gives
\begin{equation}
q_I
=
\sum_{i\in I}q_i,
\qquad
\theta_I
=
2\pi(1-q_I).
\label{eq:cluster-opening-angle}
\end{equation}
If \(q_I<1\), the fused source is again an ordinary hyperbolic cone point,
with deficit \(2\pi q_I\) and positive opening angle \(\theta_I\).

At \(q_I=1\), the opening angle vanishes. This configuration is not an
interior point of the cone-surface moduli space, but it is included in its
compactification as the limiting cuspidal channel. The compactified
collision condition is therefore
\begin{equation}
q_I
=
\sum_{i\in I}q_i
\leq1.
\label{eq:admissible-collision}
\end{equation}
The strict inequality describes a conical collision, while equality gives
the cusp limit
\cite{Eberhardt:2023rzz}.

There is a minor holonomy subtlety at the endpoint. In the lift used in
Section~\ref{subsec:defect-sources}, the local monodromy is represented by
\begin{equation}
U_I
\sim
\exp(-2\pi q_IJ).
\label{eq:cluster-elliptic-monodromy}
\end{equation}
As \(q_I\to1\), this approaches a central element of
\(\mathrm{SL}(2,\mathbb R)\), and hence the identity in
\(\mathrm{PSL}(2,\mathbb R)\). The holonomy of a complete hyperbolic cusp is
instead parabolic. The cusp is reached through a singular geometric limit
in which the elliptic fixed point moves to the ideal boundary; it is not
obtained by simply equating a finite elliptic representative with a
parabolic element
\cite{Witten:2020wvy,Eberhardt:2023rzz}.

For \(q_I>1\), the formal opening angle is negative. Such a source does not
describe an ordinary hyperbolic cone point on the gravitational integration
cycle used here and is excluded from the localization locus. We do not
consider whether it might admit an interpretation on another BF contour.

Equation~\eqref{eq:admissible-collision} identifies the possible collision
channels but not their quantum coefficients. The local BF equation fixes the
parameter \(q_I\) of the fused source, while the compactified gravitational
geometry determines the measure carried by the collision locus. The
corresponding contact coefficients require the conical Weil--Petersson
measure and are not fixed by source addition alone
\cite{Turiaci:2020fjj,Eberhardt:2023rzz}.

For identical defects of fractional deficit \(q\), an \(r\)-fold collision
is allowed in the compactified cone locus when
\begin{equation}
rq\leq1.
\label{eq:identical-collision-threshold}
\end{equation}
It produces an ordinary cone point for \(rq<1\) and reaches the cusp for
\(rq=1\). In particular, defects with \(q>\tfrac12\) admit no pairwise
collision, while defects with \(0<q<\tfrac12\) may collide pairwise to form
another cone point. At the wall \(q=\tfrac12\), a pair of identical defects
can meet only in the cuspidal limit. In our convention this wall is included
in the sharp range, with the cusp understood by the limiting prescription.
These thresholds agree with the chamber structure of the general-angle
defect expansion
\cite{Turiaci:2020fjj}.

For nonidentical defects, different subsets may satisfy different collision
conditions. For example,
\begin{equation}
q_1+q_2\leq1,
\qquad
q_1+q_2+q_3>1
\label{eq:pair-but-not-triple}
\end{equation}
allows the first two markings to coincide but excludes the triple
collision. The admissible subsets are precisely those for which
\begin{equation}
\sum_{i\in I}q_i\leq1.
\label{eq:weighted-subset-condition}
\end{equation}

This is also the coincidence criterion for marked points with Hassett
weights \(q_i\)
\cite{Hassett:2003}. The agreement of the inequalities is a local result; it
does not by itself identify the complete localization space with a Hassett
compactification. The global identification additionally requires the
stability condition on every component and the treatment of nodal and
collision strata. For hyperbolic cone surfaces, these compactifications and
their chamber structure were established in
\cite{Anagnostou:2022cmq,Anagnostou:2023jiy}. We use those results in the
next subsection to complete the geometric interpretation of the BF
collision rule.

\subsection{Hassett stability and collision strata}
\label{subsec:weighted}

The local BF analysis gives a simple condition for a collection of moving
defects to collide. If \(I\) labels the defects in the cluster, their
fractional deficits add,
\begin{equation}
q_I
=
\sum_{i\in I}q_i,
\label{eq:weighted-cluster-sum}
\end{equation}
and the collision remains in the compactified gravitational cone locus when
\begin{equation}
q_I\leq1.
\label{eq:Hassett-coincidence}
\end{equation}
This is precisely the coincidence criterion for markings of Hassett weights
\(q_i\)~\cite{Hassett:2003}. We now show that the remaining,
componentwise part of weighted stability also follows naturally from the
gravitational integration cycle.

Let \(C_v\) be a component of a limiting hyperbolic cone surface. Denote
its genus by \(g_v\), the number of cusp ends created by the degeneration
by \(c_v\), and the conical markings carried by it by \(p_i\). In the
corresponding stable curve, these cusp ends are the branches of the nodes
incident on \(C_v\). For clarity, fixed asymptotic geodesic boundaries are
suppressed below. If \(C_v\) carries \(s_v\) such boundaries, they enter
the Gauss--Bonnet and stability conditions in the usual way, through the
replacement \(c_v\to c_v+s_v\) in the corresponding topological counting.

To apply Gauss--Bonnet, it is useful to recall the relation between the
curvature conventions. In two dimensions,
\begin{equation}
R=2K,
\end{equation}
where \(R\) is the scalar curvature and \(K\) is the Gaussian curvature.
Thus the scalar-curvature source
\begin{equation}
(R+2)\,\mathrm{vol}
=
4\pi
\sum_{p_i\in C_v}
q_i\,\delta_{p_i}
\end{equation}
is equivalent to
\begin{equation}
(K+1)\,\mathrm{vol}
=
2\pi
\sum_{p_i\in C_v}
q_i\,\delta_{p_i}.
\label{eq:Gaussian-curvature-source-component}
\end{equation}
Each conical marking therefore contributes \(2\pi q_i\) to the integrated
Gaussian curvature.

A cusp may be treated in either of two equivalent ways. It is the
zero-opening-angle limit of a cone point, corresponding to \(q\to1\), and
therefore contributes \(2\pi\) to the integrated curvature defect.
Alternatively, one may truncate the cusp along a small horocycle, apply
Gauss--Bonnet to the resulting compact surface with boundary, and then
remove the cutoff. In this limit, the horocycle boundary term reproduces
the contribution obtained from the zero-angle cone limit, leading to the
same area formula.

Let \(\overline C_v\) denote the compact surface obtained by filling in the
\(c_v\) cusp ends. Its Euler characteristic is
\begin{equation}
\chi(\overline C_v)
=
2-2g_v.
\end{equation}
The smooth part of \(C_v\) has \(K=-1\), so the regular curvature contributes
\begin{equation}
\int_{C_v^{\mathrm{reg}}}
K\,\mathrm{d}A
=
-\operatorname{Area}(C_v).
\end{equation}
Including the conical curvature defects and the limiting cusp
contributions, Gauss--Bonnet gives
\begin{equation}
-\operatorname{Area}(C_v)
+
2\pi
\sum_{p_i\in C_v}q_i
+
2\pi c_v
=
2\pi\chi(\overline C_v)
=
2\pi(2-2g_v).
\label{eq:component-Gauss-Bonnet}
\end{equation}
Solving for the area yields
\begin{equation}
\operatorname{Area}(C_v)
=
2\pi
\left(
2g_v-2+c_v
+
\sum_{p_i\in C_v}q_i
\right).
\label{eq:component-hyperbolic-area}
\end{equation}

The right-hand side has an immediate gravitational meaning. On
\(\Gamma_{\mathrm{grav}}\), the area form is reconstructed from the zweibein,
\begin{equation}
\mathrm{vol}_{C_v}
=
e^1\wedge e^2
=
\sqrt{g}\,\mathrm{d}^2x.
\label{eq:component-area-zweibein}
\end{equation}
The gravitational integration cycle contains real, orientation-preserving, nondegenerate zweibeins. A component that survives
as a two-dimensional part of the localization locus must therefore have
strictly positive area. Equation~\eqref{eq:component-hyperbolic-area} then
implies
\begin{equation}
2g_v-2+c_v
+
\sum_{p_i\in C_v}q_i
>
0.
\label{eq:global-weighted-stability}
\end{equation}
In the language of algebraic geometry, \(c_v\) counts the nodal branches
incident on \(C_v\), and
\begin{equation}
2g_v-2+c_v+\sum_{p_i\in C_v}q_i
=
\deg\!\left[
\left.
\omega_C\!\left(\sum_i q_i p_i\right)
\right|_{C_v}
\right].
\label{eq:weighted-log-canonical-degree}
\end{equation}
Hence \eqref{eq:global-weighted-stability} is precisely the condition that
the weighted log canonical bundle
\begin{equation}
\omega_C\!\left(\sum_i q_i p_i\right)
\label{eq:weighted-log-canonical-bundle}
\end{equation}
have positive degree on every irreducible component, and therefore be ample.
Together with the coincidence condition
\begin{equation}
\sum_{i\in I}q_i\leq1
\end{equation}
for coincident marked points, this is the Hassett stability condition
\cite{Hassett:2003}.

The physical interpretation is straightforward. A positive value in
\eqref{eq:global-weighted-stability} gives a component of positive
hyperbolic area and hence permits a nondegenerate zweibein. At equality, its
area vanishes and the component collapses. A negative value would require
negative area and cannot occur on \(\Gamma_{\mathrm{grav}}\).

This becomes especially transparent for a rational tail \(C_I\) carrying
the markings indexed by \(I\). Such a tail has genus zero and one cusp end
where it attaches to the rest of the limiting surface,
\begin{equation}
g(C_I)=0,
\qquad
c_I=1.
\label{eq:rational-tail-data}
\end{equation}
Its area is therefore
\begin{align}
\operatorname{Area}(C_I)
&=
2\pi
\left(
-2+1+\sum_{i\in I}q_i
\right)
\nonumber\\
&=
2\pi(q_I-1).
\label{eq:rational-tail-area}
\end{align}
The tail can remain as a separate nondegenerate hyperbolic component only
when
\begin{equation}
q_I>1.
\label{eq:stable-rational-tail}
\end{equation}
At \(q_I=1\), its area vanishes and the family reaches the contraction wall.
For \(q_I<1\), the formal area is negative, so no such component belongs to
the gravitational integration cycle. In the compactified moduli problem, a
tail with
\begin{equation}
q_I\leq1
\label{eq:unstable-rational-tail}
\end{equation}
is consequently collapsed, and its markings become a coincident
configuration on the component that remains.

The BF description therefore recovers both ingredients of weighted
stability. A cluster of markings may coincide when its total weight is at
most one. Independently, Gauss--Bonnet and the nondegeneracy of the zweibein
require every surviving component to satisfy
\eqref{eq:global-weighted-stability}. For a rational tail these are two
aspects of the same degeneration: once its total weight drops to one or
below, its area becomes nonpositive, the tail collapses, and its markings
are allowed to coincide on the remaining component.

The relation between the ordinary and weighted compactifications is
described by the Hassett reduction morphism
\begin{equation}
\rho_{\mathbf q}:
\overline{\mathcal{M}}_{g,L}
\longrightarrow
\overline{\mathcal{M}}_{g,\mathbf q},
\qquad
\mathbf q=(q_1,\ldots,q_L).
\label{eq:Hassett-reduction-map}
\end{equation}
Starting from weight one at each defect marking, the morphism
\(\rho_{\mathbf q}\) lowers the weights to \(q_i\) and contracts precisely
those rational components that become unstable after this reduction
\cite{Hassett:2003}. The numerical stability conditions derived above
coincide with the stability conditions defining the corresponding Hassett
space. Thus the proper algebraic compactification, together with its global
boundary and coincidence stratification, is supplied by
\(\overline{\mathcal{M}}_{g,\mathbf q}\). For hyperbolic cone surfaces, the
corresponding compactification was identified with the weighted moduli space
in \cite{Anagnostou:2022cmq}.

To make the contraction more explicit, let
\begin{equation}
\delta_{0,I}
\subset
\overline{\mathcal{M}}_{g,L}
\label{eq:rational-tail-boundary-divisor}
\end{equation}
denote the ordinary Deligne--Mumford boundary divisor whose generic curve
contains a genus-zero tail carrying the markings in \(I\) and attached to
the rest of the curve at a single node. If
\begin{equation}
q_I:=\sum_{i\in I}q_i\leq1,
\end{equation}
then, after the weights are reduced, this rational tail is unstable and is
contracted by \(\rho_{\mathbf q}\). Its markings are mapped to the
attachment point and therefore become coincident. Consequently,
\begin{equation}
\rho_{\mathbf q}(\delta_{0,I})
=
\Delta_I,
\qquad
\Delta_I
=
\left\{
p_i=p_j
\ \text{for all }i,j\in I
\right\}
\subset
\overline{\mathcal{M}}_{g,\mathbf q}.
\label{eq:weighted-coincidence-locus}
\end{equation}

If \(r=|I|\) markings coincide, their common position remains free, while
the remaining \(r-1\) relative positions are constrained to vanish.
Accordingly,
\begin{equation}
\operatorname{codim}_{\mathbb C}\Delta_I=r-1.
\label{eq:r-fold-diagonal-codimension}
\end{equation}
This distinction is important because the source
\(\delta_{0,I}\) is always a divisor in the ordinary
Deligne--Mumford space, whereas its image need not be a divisor in the
weighted space. For \(|I|=2\),
\begin{equation}
\Delta_{i_1i_2}
=
\{p_{i_1}=p_{i_2}\}
\label{eq:pairwise-weighted-diagonal}
\end{equation}
has complex codimension one, so both
\(\delta_{0,\{i_1,i_2\}}\) and its image are divisors. For three markings,
however,
\begin{equation}
\Delta_{i_1i_2i_3}
=
\{p_{i_1}=p_{i_2}=p_{i_3}\}
\label{eq:triple-weighted-diagonal}
\end{equation}
has complex codimension two. More generally, for \(|I|\geq3\),
\(\delta_{0,I}\) is an exceptional divisor of the reduction morphism,
contracted onto a locus of complex codimension \(|I|-1>1\).

The weights determine not only which coincidence loci are allowed, but
also how these loci can intersect. Let
\(I\cap J\neq\varnothing\). Imposing simultaneously the conditions defining
\(\Delta_I\) and \(\Delta_J\) forces all markings in \(I\cup J\) to
coincide. The two strata can therefore intersect only when the larger
coincidence is also admissible:
\begin{equation}
q_{I\cup J}
=
\sum_{i\in I\cup J}q_i
\leq
1.
\label{eq:overlapping-diagonal-condition}
\end{equation}
Thus the admissible coincidence loci cannot be treated independently.
Their incidence relations are constrained by the admissibility of the
larger clusters formed at their intersections. These relations are part of
the global weighted stratification of the Hassett space.

This gives a direct geometric interpretation of the local BF fusion rule.
For sources reduced to a common elliptic direction, the coincident source
depends only on the total weight
\begin{equation}
q_I=\sum_{i\in I}q_i.
\end{equation}
The local BF theory therefore determines the effective conical defect
produced by the collision, but the single number \(q_I\) does not determine
the global geometry of the corresponding coincidence locus. Its
codimension, incidence relations, multiplicities, and normal geometry
depend on the embedding
\(\Delta_I\subset\overline{\mathcal M}_{g,\mathbf q}\).

The same weighted condition follows from the hyperbolic geometry. The
fractional deficits of aligned sources add under collision, while
Gauss--Bonnet relates the resulting total deficit to the area of each
component. Requiring a nondegenerate zweibein, and hence positive
hyperbolic area on every component, reproduces the weighted stability
condition. When a rational tail ceases to satisfy this condition after the
weights are reduced, its hyperbolic description degenerates and the
Hassett reduction contracts the component, leaving the coincident defect
markings on the remaining curve.

The elementary local defect kernel is unchanged as the weights vary across
the Hassett chambers. What changes is the collection of coincidence strata
that are globally admissible. In the open sharp chamber,
\begin{equation}
q_i>\frac12
\qquad
\text{for every }i,
\label{eq:strictly-sharp-weights}
\end{equation}
one has
\begin{equation}
q_i+q_j>1
\end{equation}
for every pair of defect markings. No pairwise collision is therefore
allowed, and hence no collision involving two or more defect markings can
occur. In this chamber, the weighted reduction introduces no new
coincidence loci associated with defect collisions.

At the marginal value
\begin{equation}
q_i=q_j=\frac12,
\end{equation}
the pair lies exactly on the wall
\begin{equation}
q_i+q_j=1.
\end{equation}
The coincidence \(p_i=p_j\) is then allowed in the weighted moduli space.
On the hyperbolic side, the effective defect has total fractional deficit
one, so its opening angle vanishes and the cone reaches the cusp limit
described in Section~\ref{subsec:admissible}. This statement should not be
confused with a claim that the cusp limit is itself a boundary divisor of
the Hassett space: the locus \(\Delta_{ij}\) is a weighted coincidence
divisor, while ``cusp endpoint'' refers to the degeneration of the
effective hyperbolic cone angle.

For blunter defects, subsets with
\begin{equation}
q_I<1
\end{equation}
may collide to form an ordinary conical cluster, whereas
\begin{equation}
q_I=1
\end{equation}
gives the cusp limit. The walls separating the Hassett chambers are
\begin{equation}
\sum_{i\in I}q_i=1.
\end{equation}
More precisely, as the weights are decreased across such a wall from
\begin{equation}
q_I>1
\qquad\text{to}\qquad
q_I\leq1,
\end{equation}
the coincidence of the markings in \(I\) becomes admissible. Correspondingly,
the reduction morphism changes by contracting the rational tail
\(\delta_{0,I}\) onto the weighted diagonal \(\Delta_I\). The chamber
structure and wall-crossing of the cone-surface volume polynomials were
analyzed in \cite{Anagnostou:2023jiy}. Related collision and contact-term
effects in JT gravity with general conical defects appear in
\cite{Turiaci:2020fjj,Eberhardt:2023rzz,Lin:2023trc,
Kruthoff:2024rpx}.

\subsection{Defect correlators and their generating functional}
\label{subsec:correlation-functions}

For a genus-\(g\) surface with \(n\) asymptotic boundaries and \(L\)
integrated defects, we define
\begin{equation}
\left\langle
\prod_{i=1}^{L}\mathfrak{D}_{q_i}
\right\rangle_{g,n}
=
\frac{1}{\operatorname{Vol}\mathcal{G}}
\int_{\Gamma_{\mathrm{grav}}}
\mathcal{D}A\,\mathcal{D}X\,
e^{-I_{\mathrm{BF}}[A,X]}
\prod_{i=1}^{L}
\left[
\int_{\Sigma}
\Omega_{n_i}(A;p_i)\,
e^{2\pi iq_i\langle n_i(p_i),X(p_i)\rangle}
\right].
\label{eq:section4-defect-correlator}
\end{equation}
Here \(\mathcal G\) is the gauge group compatible with the prescribed
asymptotic boundary conditions. The integrations over the positions \(p_i\)
are part of the operators \(\mathfrak D_{q_i}\); no additional
marked-point measure is introduced.

Using
\begin{equation}
\int_{\Sigma}
\left\langle
X,n_i
\right\rangle
\delta_{p_i}
=
\left\langle
X(p_i),n_i(p_i)
\right\rangle,
\label{eq:delta-form-evaluation}
\end{equation}
the \(X\)-dependent terms combine into
\begin{equation}
\exp\!\left[
i\int_{\Sigma}
\left\langle
X,
F_A
+
2\pi\sum_{i=1}^{L}
q_i n_i\,\delta_{p_i}
\right\rangle
\right].
\label{eq:multi-defect-X-exponent}
\end{equation}
Integration over \(X\) therefore imposes
\begin{equation}
F_A
+
2\pi\sum_{i=1}^{L}
q_i n_i\,\delta_{p_i}
=
0.
\label{eq:multi-defect-localization-constraint}
\end{equation}
The connection is flat away from the insertions and has the prescribed
elliptic conjugacy class around each marked point.

For fixed positions \(\mathbf p=(p_1,\ldots,p_L)\), it is useful to package the sourced-flatness constraint in the
moment-map-like expression
\begin{equation}
\mu_{\mathbf q}(A;\mathbf p)
=
F_A
+
2\pi\sum_{i=1}^{L}
q_i n_i\,\delta_{p_i},
\qquad
\mathbf q=(q_1,\ldots,q_L).
\label{eq:defect-moment-map}
\end{equation}
The \(X\) integration restricts the connection to
\(\mu_{\mathbf q}^{-1}(0)\). Quotienting by internal gauge transformations
gives, for fixed \(\mathbf p\), the moduli space of flat connections with
the prescribed elliptic conjugacy classes. The positions themselves are
integrated with the two-forms \(\Omega_{n_i}(A;p_i)\). On the gravitational
cycle, the diffeomorphism redundancy identifies configurations related by
motions of the marked points, leaving their relative positions as moduli.

On a smooth stratum, the gauge-field zero modes inherit the
Atiyah--Bott--Goldman symplectic form
\begin{equation}
\Omega_{\mathcal A}
\left(
\delta_1A,\delta_2A
\right)
=
\int_{\Sigma}
\left\langle
\delta_1A\wedge\delta_2A
\right\rangle.
\label{eq:Atiyah-Bott-form-defects}
\end{equation}
After imposing the moment-map constraint and dividing by gauge
transformations, the Atiyah--Bott--Goldman form descends to a symplectic
form \(\omega_{\mathrm{red},\mathbf q}\) on the reduced moduli space. On a
smooth component of complex dimension \(d_{g,n,L}\), the corresponding
Liouville measure is
\begin{equation}
\mathrm d\mu_{\mathrm{red},\mathbf q}
=
\frac{
\omega_{\mathrm{red},\mathbf q}^{\,d_{g,n,L}}
}{
d_{g,n,L}!
}.
\end{equation}

The gravitational integration cycle selects the Fuchsian component on which
the flat connection determines a nondegenerate hyperbolic cone metric. On
this component, the reduced Atiyah--Bott--Goldman form agrees, in the
normalization adopted here, with the conical Weil--Petersson form
\cite{Schumacher:2005aa,Schumacher:2008ff,Wolpert:1983aa}:
\begin{equation}
\left.
\omega_{\mathrm{red},\mathbf q}
\right|_{\Gamma_{\mathrm{grav}}}
=
\omega_{\mathrm{WP},\mathbf q}.
\end{equation}
We derive this reduced measure and fix its normalization in
Appendix~\ref{app:moment_map_reduction}.. For fixed geodesic boundary lengths, the complex dimension of the stable
cone-surface moduli space is
\begin{equation}
d_{g,n,L}
=
3g-3+n+L.
\label{eq:cone-moduli-complex-dimension}
\end{equation}
The one-defect disk is exceptional: its position orbit is removed completely
by the residual \(\mathrm{PSL}(2,\mathbb R)\) symmetry, as discussed in
Section~\ref{subsec:fixing-defect-position}. We henceforth restrict this
moduli-space description to stable configurations.

On the open locus of distinct marked points, the gravitational correlator
takes the form
\begin{equation}
\left\langle
\prod_{i=1}^{L}\mathfrak D_{q_i}
\right\rangle_{g,n}^{\mathrm{sep}}
=
\int_{\mathcal M^{\mathrm{cone}}_{g,n;\mathbf q}}
\frac{
\omega_{\mathrm{WP},\mathbf q}^{\,d_{g,n,L}}
}{
d_{g,n,L}!
}
\,
\mathcal B_{g,n},
\label{eq:separated-defect-moduli-integral}
\end{equation}
where \(\mathcal B_{g,n}\) denotes the factors associated with the
asymptotic boundaries. Equivalently, one may remove the asymptotic trumpet
regions, perform the cone-surface moduli integral at fixed geodesic boundary lengths, and sew the trumpets back to the compact core. The explicit
boundary factors are restored in Section~\ref{sec:partition}.

The open moduli space describes separated cone points and does not include
their collision limits. If the defects in a subset \(I\) approach one
another, the BF source-composition rule gives
\begin{equation}
q_I
=
\sum_{i\in I}q_i,
\qquad
a_I
=
1-q_I.
\label{eq:section4-cluster-parameter}
\end{equation}
Restriction to the gravitational sector determines whether the fused
source is admissible, while positivity of the hyperbolic area of each
component gives the stability condition for its degenerations. Together,
these conditions agree with those defining the weighted compactification
\(\overline{\mathcal M}_{g,n;\mathbf q}\), described in
Section~\ref{subsec:weighted}.

On the open locus of separated cone points, the BF measure reduces to the
conical Weil--Petersson measure. Passing to
\(\overline{\mathcal M}_{g,n;\mathbf q}\) does not change this local
measure; it adds the admissible collision and degeneration strata and
organizes their intersections. In this intersection-theoretic sense, the
full correlator has the schematic form
\begin{equation}
\left\langle
\prod_{i=1}^{L}\mathfrak D_{q_i}
\right\rangle_{g,n}
=
\int_{\overline{\mathcal M}_{g,n;\mathbf q}}
\mathrm d\mu_{\mathrm{WP},\mathbf q}\,
\mathcal B_{g,n}.
\label{eq:weighted-defect-moduli-integral}
\end{equation}
Here \(\mathcal B_{g,n}\) denotes the remaining boundary and sewing
factors.

For the genus-zero disk, this distinction can be made explicit. A single
defect has no relative-position modulus and reduces, after the residual
position quotient, to the elementary elliptic kernel of
Section~\ref{sec:quantization}. With several defects, relative positions
remain and the weighted compactification contains coincidence strata
whenever the corresponding fused sources are gravitationally admissible.
In a fully coincident sector, the fused BF source determines the local
fixed-point block, while the weighted moduli-space geometry determines how
that block enters the complete amplitude. These contributions are evaluated
in Section~\ref{sec:partition}.

To organize different defect species, introduce a fugacity density
\(\lambda(q)\) and define
\begin{equation}
Z_{g,n}[\lambda]
=
\left\langle
\exp\!\left[
\int_0^1
\mathrm dq\,
\lambda(q)\mathfrak D_q
\right]
\right\rangle_{g,n}.
\label{eq:defect-generating-functional}
\end{equation}
Expanding the exponential gives
\begin{align}
Z_{g,n}[\lambda]
&=
\sum_{L=0}^{\infty}
\frac{1}{L!}
\int_0^1
\prod_{i=1}^{L}
\left[
\mathrm dq_i\,\lambda(q_i)
\right]
\left\langle
\prod_{i=1}^{L}
\mathfrak D_{q_i}
\right\rangle_{g,n}.
\label{eq:defect-generating-expansion}
\end{align}

For a discrete set of defect species,
\begin{equation}
\lambda(q)
=
\sum_{\alpha}
\lambda_\alpha\,
\delta(q-q_\alpha),
\label{eq:discrete-defect-species}
\end{equation}
this becomes
\begin{equation}
Z_{g,n}[\boldsymbol\lambda]
=
\sum_{\{L_\alpha\}}
\prod_\alpha
\frac{
\lambda_\alpha^{L_\alpha}
}{
L_\alpha!
}
\left\langle
\prod_\alpha
\mathfrak D_{q_\alpha}^{\,L_\alpha}
\right\rangle_{g,n}.
\label{eq:discrete-generating-expansion}
\end{equation}
The factors \(1/L_\alpha!\) are the usual symmetry factors for identical
defects.

\subsection{BF origin of cotangent-line and descendant classes}
\label{subsec:BF-psi-classes}

Fixing the position of an elliptic defect leaves a compact
\(U(1)\) stabilizer. On the gravitational cycle, this subgroup
rotates an orthonormal frame at the marked point. As the marked surface
varies over moduli space, the stabilizers form a principal
\(U(1)\) bundle. We will identify the line bundle associated with
its cotangent representation with the tautological cotangent line
\(\mathcal L_i\), whose first Chern class is
\(\psi_i=c_1(\mathcal L_i)\).

For the BF defect observable introduced above, fix the position of the
\(i\)-th defect and choose a local gauge in which
\begin{equation}
n_i
=
J.
\label{eq:elliptic-source-gauge}
\end{equation}
The subgroup preserving this representative is the compact elliptic
stabilizer
\begin{equation}
\operatorname{Stab}(J)
=
\left\{
e^{\varphi J}
\right\}
\simeq
U(1).
\label{eq:elliptic-stabilizer-U1}
\end{equation}
To identify its geometric action, use the gravitational decomposition
\begin{equation}
A
=
-\omega J+e^aP_a.
\label{eq:gravitational-decomposition-psi}
\end{equation}
The adjoint action of the stabilizer on the translational generators is
\begin{equation}
\operatorname{Ad}_{e^{\varphi J}}P_a
=
R_a{}^b(\varphi)P_b,
\qquad
R(\varphi)\in SO(2).
\label{eq:U1-rotation-translations}
\end{equation}

On the nondegenerate gravitational locus, the zweibein identifies the
internal translational plane with the tangent plane:
\begin{equation}
e_p:
T_p\Sigma
\overset{\sim}{\longrightarrow}
\operatorname{span}\{P_1,P_2\},
\qquad
v^\mu\partial_\mu
\longmapsto
e^a{}_\mu v^\mu P_a.
\label{eq:zweibein-map}
\end{equation}
Equivalently, the translational generators \(P_a\) provide a basis of the
internal two-dimensional space \(\mathbb R^2_{\rm int}\), while the inverse
zweibein identifies this internal space with the tangent space,
\begin{equation}
e_p^{-1}:\mathbb R^2_{\rm int}\longrightarrow T_p\Sigma.
\end{equation}
In particular, the basis vector \(P_a\) is mapped to
\begin{equation}
E_a\equiv e_p^{-1}(P_a)
      =e_a{}^\mu\partial_\mu.
\label{eq:zweibein-basis-identification}
\end{equation}
These vectors satisfy
\begin{equation}
e^a(E_b)=\delta^a{}_b,
\end{equation}
and, since \(g=\delta_{ab}e^a\otimes e^b\),
\begin{equation}
g(E_a,E_b)=\delta_{ab}.
\end{equation}
Thus \(E_a\) is the orthonormal tangent frame corresponding to the internal
basis \(P_a\). Under this identification, an internal rotation
\(P_a\mapsto R_a{}^b(\varphi)P_b\) induces the same \(SO(2)\) rotation of
the tangent frame,
\begin{equation}
E_a\mapsto R_a{}^b(\varphi)E_b.
\end{equation}
Thus the internal \(U(1)\simeq SO(2)\) stabilizer is
identified with rotations of the local oriented orthonormal frame. At a
conical marking, this identification is understood through the limiting
frame in a punctured neighborhood of the defect.

The complementary directions of the gauge orbit move the marked point.
Indeed, an infinitesimal diffeomorphism generated by \(\xi\) satisfies
\begin{equation}
\mathcal L_\xi A
=
D_A(\iota_\xi A)+\iota_\xi F_A.
\label{eq:gauge-diffeomorphism-identity}
\end{equation}
Away from the localized source, \(F_A=0\), and the transformation with
gauge parameter \(\iota_\xi A\) agrees with the corresponding
diffeomorphism. Through the nondegenerate zweibein, the two translational
directions therefore become the two tangent directions that move the
marking. After these position modes are quotiented, the compact rotational
stabilizer remains.

As the marked surface varies over moduli space, these residual circles
assemble into a principal bundle
\begin{equation}
U(1)
\longrightarrow
\mathcal P_i
\longrightarrow
\mathcal M_{g,n+L}.
\label{eq:defect-frame-principal-bundle}
\end{equation}
A point of \(\mathcal P_i\) consists of a gravitational configuration
together with a choice of unit tangent direction at the \(i\)-th marking,
or equivalently an oriented orthonormal frame there. Different local
choices are related on overlaps by \(U(1)\)-valued transition
functions.

 To obtain the tautological cotangent line rather than the tangent line,
we use the dual one-dimensional representation of the residual
\(U(1)\).  Let \(\chi_{\mathrm{cot}}\) denote the representation with
which an oriented complex cotangent covector transforms under a frame
rotation.  If the complex tangent frame transforms as
\begin{equation}
E_1+iE_2
\longmapsto
e^{i\varphi}(E_1+iE_2),
\end{equation}
then the dual complex cotangent covector transforms with the opposite
weight,
\begin{equation}
\vartheta
\longmapsto
e^{-i\varphi}\vartheta,
\qquad
\chi_{\mathrm{cot}}(e^{i\varphi})=e^{-i\varphi}.
\label{eq:cotangent-U1-character}
\end{equation}
The corresponding line bundle associated to the residual frame bundle
\(\mathcal P_i\) is therefore
\begin{equation}
\mathbb L_i^{\mathrm{BF}}
=
\mathcal P_i
\times_{U(1),\chi_{\mathrm{cot}}}
\mathbb C.
\label{eq:BF-associated-cotangent-line}
\end{equation}
On the gravitational locus this is canonically identified with the usual
cotangent line at the marked point,
\begin{equation}
\mathbb L_i^{\mathrm{BF}}
\simeq
\mathbb L_i,
\qquad
(\mathbb L_i)_{[\Sigma,p_1,\ldots,p_{n+L}]}
=
T_{p_i}^{*\,1,0}\Sigma.
\label{eq:BF-cotangent-line-identification}
\end{equation}
Consequently,
\begin{equation}
\psi_i
=
c_1(\mathbb L_i)
=
c_1(\mathbb L_i^{\mathrm{BF}}).
\label{eq:psi-as-BF-Chern-class}
\end{equation}

This construction also explains why retaining only the compact stabilizer
does not lose any information relevant to \(\psi_i\). A nonzero vector in
a complex line has the decomposition
\begin{equation}
\mathbb C^\times
\simeq
\mathbb R_{>0}\times U(1).
\label{eq:Cstar-radial-angular}
\end{equation}
The positive radial factor is contractible. The topology of a complex line
bundle, and in particular its first Chern class, is therefore completely
captured by the unit-circle bundle. The residual BF \(U(1)\) is precisely
this compact topological part.

The same statement can be formulated intrinsically using the universal
curve
\begin{equation}
\pi:
\mathcal C_{g,n+L}
\longrightarrow
\mathcal M_{g,n+L},
\label{eq:universal-curve-BF}
\end{equation}
with section
\begin{equation}
s_i:
\mathcal M_{g,n+L}
\longrightarrow
\mathcal C_{g,n+L}.
\label{eq:marked-section-BF}
\end{equation}
The tautological cotangent line is
\begin{equation}
\mathcal L_i
=
s_i^*\omega_\pi,
\label{eq:tautological-line-universal-curve}
\end{equation}
where \(\omega_\pi\) is the relative cotangent bundle. The BF construction
identifies its unit-circle frame bundle with the residual rotational bundle
of the defect. The class \(\psi_i=c_1(\mathcal L_i)\) therefore measures the
global twisting of the local frame over moduli space; it is a characteristic
class on the reduced configuration space, not a local spacetime operator
inserted at \(p_i\).

We next explain how the same structure controls the dependence on the
defect parameter. The fixed-position operator \(\mathcal O_{q_i}(p_i)\) should not be
identified with a descendant insertion. Its parameter fixes the elliptic
conjugacy class, or equivalently the level at which the gauge-theoretic
symplectic reduction is performed.  Changing the cone angle therefore
changes the reduction level associated with the residual \(U(1)\).

By the standard variation formula for Hamiltonian symplectic reduction,
the cohomology class of the reduced symplectic form varies with the
reduction level in the direction of the first Chern class of the
corresponding stabilizer bundle. In the present case this gives
schematically
\begin{equation}
\frac{\partial}{\partial \mu_i}
[\omega_{\mathrm{red}}]
\propto
c_1(\mathbb L_i)
=
\psi_i,
\label{eq:DH-variation-psi}
\end{equation}
where \(\mu_i\) denotes an appropriate parameter for the elliptic reduction
level. The BF argument therefore identifies the cohomology class that must control the defect-parameter dependence. It does not by itself fix the precise normalization of that dependence.

The Weil--Petersson geometry of hyperbolic cone metrics can be defined
intrinsically from their variation in holomorphic families
\cite{Schumacher:2005aa,Schumacher:2008ff}. In a chamber without additional
collision corrections, its cohomology class agrees with the analytic
continuation of the bordered Weil--Petersson class.

For geodesic boundaries of lengths \(b_i\), the cohomology class is
\cite{Mirzakhani:2006,DoNorbury:2009}
\begin{equation}
[\omega_{\mathrm{WP}}(\mathbf b)]
=
2\pi^2\kappa_1
+
\frac12
\sum_i b_i^2\psi_i.
\label{eq:WP-boundary-class-BF}
\end{equation}
Analytically continuing
\begin{equation}
b_i=i\theta_i,
\qquad
\theta_i=2\pi(1-q_i),
\label{eq:boundary-cone-continuation-BF}
\end{equation}
gives, in a chamber in which no additional collision corrections are
present,
\begin{equation}
[\omega_{\mathrm{WP}}(\boldsymbol\theta)]
=
2\pi^2\kappa_1
-
\frac12
\sum_i\theta_i^2\psi_i.
\label{eq:conical-WP-class-BF}
\end{equation}
Thus the BF reduction explains why the cotangent-line class \(\psi_i\)
governs the dependence on the elliptic defect parameter, while the
Weil--Petersson normalization fixes its coefficient.

The microscopic BF insertion and the descendant factor arise at different
stages of the reduction. The fixed-position operator
\(\mathcal O_{q_i}(p_i)\) is inserted in the spacetime path integral and
fixes the local elliptic source. After localization and symplectic
reduction, the conical Weil--Petersson class contains the term
\(-\tfrac12\theta_i^2\psi_i\), so the corresponding contribution to the
reduced volume form is
\begin{equation}
\exp\!\left(
-\frac{\theta_i^2}{2}\psi_i
\right)
=
\sum_{d_i=0}^{\infty}
\frac{1}{d_i!}
\left(
-\frac{\theta_i^2}{2}
\right)^{d_i}
\psi_i^{d_i}.
\label{eq:descendant-generating-factor-BF}
\end{equation}
The coefficients of the opening-angle expansion therefore give
intersection numbers containing the descendant classes
\(\psi_i^{d_i}\). This is a statement about the reduced moduli-space
integrand, not an operator identity or a Taylor expansion of the microscopic
BF insertion in powers of \(X\).

For several separated defects, the same argument gives
\begin{equation}
\exp\!\left[
-\frac12
\sum_{i=1}^{L}\theta_i^2\psi_i
\right]
=
\prod_{i=1}^{L}
\sum_{d_i\geq0}
\frac{1}{d_i!}
\left(
-\frac{\theta_i^2}{2}
\right)^{d_i}
\psi_i^{d_i}.
\label{eq:multi-defect-descendant-generating-factor}
\end{equation}

The preceding discussion applies directly on the separated-defect locus.
When weighted collisions are allowed, the global geometry must be treated
more carefully. If
\begin{equation}
q_I=\sum_{i\in I}q_i\leq1,
\end{equation}
the markings in \(I\) may coincide in the Hassett compactification. The ordinary Deligne--Mumford rational-tail divisor $\delta_{0,I}$ is then contracted by the Hassett reduction morphism. Writing
\[
\mathbf q=(q_1,\ldots,q_L),
\]
we denote the corresponding weighted moduli space by
$\overline{\mathcal M}_{g,(1^l,\mathbf q)}$.  Here the first $l$ weight-one markings label the geodesic boundary components at the level of the underlying pointed curve; their lengths are retained separately as Weil--Petersson parameters.  The entries of
$\mathbf q$ are the Hassett weights of the defect markings.  The
reduction morphism is
\begin{equation}
\rho_{\mathbf q}:
\overline{\mathcal M}_{g,l+L}
\longrightarrow
\overline{\mathcal M}_{g,(1^l,\mathbf q)} .
\label{eq:Hassett-reduction-descendants}
\end{equation}
For every subset $I$ satisfying
$q_I=\sum_{i\in I}q_i\leq1$, the divisor $\delta_{0,I}$ is contracted
onto the corresponding weighted coincidence locus $\Delta_I$.

Accordingly, one should not interpret the conical Weil--Petersson form
itself as simply acquiring delta-function support on the diagonals.
Rather, comparison of the relevant cohomology classes across the reduction
morphism produces corrections involving the contracted boundary strata.
Schematically,
\begin{equation}
\rho_{\mathbf q}^*
[\omega_{\mathrm{WP},\mathbf q}]
=
[\omega_{\mathrm{WP}}^{\mathrm{sep}}]
+
\sum_I
C_I(\mathbf q)\,
[\delta_{0,I}]
+\cdots,
\label{eq:weighted-WP-pullback-corrections}
\end{equation}
with the precise coefficients determined by the weighted geometry.
After contraction, these terms encode the contribution of the admissible
collision strata in the weighted moduli space.

From the BF viewpoint, the local origin of these strata is again simple:
when several aligned defects collide, their sources add,
\begin{equation}
q_I=\sum_{i\in I}q_i.
\end{equation}
The BF source rule therefore determines the effective local defect, while
the Hassett reduction determines how the corresponding coincidence locus
is embedded globally in the compactified moduli space. The local BF data
alone do not determine its codimension, incidence relations, or normal
geometry.

\subsection{Weighted collision geometry and string equations}
\label{subsec:defect-descendants}

The preceding subsection identified the residual rotational bundle of a
moving BF defect with the tautological cotangent-line bundle. We now turn
to the additional geometry produced by collisions of several integrated
defects.

Recall the parametrization
\begin{equation}
\theta_i
=
2\pi a_i,
\qquad
a_i
=
1-q_i,
\label{eq:descendant-angle-parameters}
\end{equation}
where \(a_i=0\) is the cusp limit and \(a_i=1\) corresponds to a smooth
marked point. On the locus of separated defects, the contribution of the
\(i\)-th marking to the normalized Weil--Petersson class is
\begin{equation}
-a_i^2\psi_i.
\label{eq:individual-defect-WP-response}
\end{equation}

When the defects in a subset \(I\) collide, the BF source-composition rule
gives
\begin{equation}
q_I
=
\sum_{i\in I}q_i,
\qquad
a_I
=
1-q_I.
\label{eq:descendant-cluster-parameters}
\end{equation}
As shown in Section~\ref{subsec:admissible}, the fused source remains in
the gravitational cone sector, including its cuspidal boundary, precisely
when
\begin{equation}
q_I
\leq
1,
\qquad\text{equivalently}\qquad
a_I
\geq
0.
\label{eq:descendant-admissibility}
\end{equation}
For \(q_I<1\), the fused source is another cone point; at \(q_I=1\), it
reaches the cusp limit.

This is also the coincidence condition for Hassett weights \(q_i\). Let $\rho_{\mathbf q}$ be the Hassett reduction morphism defined in
Eq.~\eqref{eq:Hassett-reduction-descendants}.  If \(q_I\leq1\), a rational tail
carrying the markings in \(I\) is contracted, and the divisor
\(\delta_{0,I}\) is mapped to the weighted coincidence locus.

With
\begin{equation}
\widehat\omega_{\mathrm{WP}}
\equiv
\frac{\omega_{\mathrm{WP}}}{2\pi^2},
\label{eq:normalized-WP-class}
\end{equation}
we use the conical Weil--Petersson class derived in Eq.~(1.2) of~\cite{Eberhardt:2023rzz} and further discussed in Sections~2.1--2.2 of that reference. Its interpretation under the reduction morphism is provided by the geometry of weighted pointed curves
\cite{Hassett:2003,Anagnostou:2022cmq,Anagnostou:2023jiy}.
In the notation adopted here, its pullback to the ordinary
Deligne--Mumford compactification is
\begin{equation}
\rho_{\mathbf q}^{*}
\bigl[\widehat\omega_{\mathrm{WP}}^{\mathrm{wt}}\bigr]
=
\kappa_1
-\sum_{i=1}^{L}a_i^2\psi_i
+\frac{1}{4\pi^2}
\sum_{r=1}^{n}b_r^2\psi_r^{\partial}
+
\sum_{\substack{
I\subset\{1,\ldots,L\}\\
|I|\geq2,\;q_I\leq1
}}
a_I^2\delta_{0,I}.
\label{eq:weighted-WP-pullback}
\end{equation}
The first three terms give the class on the locus of separated markings.
The final sum is supported on rational-tail divisors contracted by
\(\rho_{\mathbf q}\). Such a divisor occurs precisely when the markings in
\(I\) are allowed to coincide, \(q_I\leq1\), and its coefficient is fixed
by the effective opening fraction $a_I=1-q_I$.

The BF source equation determines the fused parameter
\(q_I=\sum_{i\in I}q_i\), and the gravitational cycle selects the admissible
channels \(q_I\leq1\). The coefficients of the divisor terms in
\eqref{eq:weighted-WP-pullback} are instead fixed by the global
Weil--Petersson geometry of the weighted compactification.

\subsubsection{The source-free defect and the string equation}
\label{subsubsec:string-equation}

The string insertion arises naturally as the source-free endpoint of the BF
defect family. Setting \(q=0\) gives
\begin{equation}
\mathfrak D^{\mathrm{BF}}_0
=
\int_\Sigma
\Omega_n(A).
\label{eq:source-free-BF-defect}
\end{equation}
Since the exponential factor is absent, this insertion does not modify the
BF curvature equation. On the gravitational integration cycle,
\begin{equation}
\left.
\mathfrak D^{\mathrm{BF}}_0
\right|_{\Gamma_{\mathrm{grav}}}
=
\int_\Sigma
\sqrt g\,\mathrm d^2x.
\label{eq:source-free-gravitational-defect}
\end{equation}
Thus the \(q=0\) endpoint is a freely moving smooth marked point with no
curvature source.

Its quantum normalization is fixed independently by the exact elementary BF
kernel. Setting \(q=0\) in the one-defect disk amplitude gives
\begin{align}
Z_{\mathfrak D_0}^{\mathrm{disk}}(\beta)
&=
\int_0^\infty
\mathrm dE\,
\frac{
\cosh(2\pi\sqrt E)
}{
2\pi\sqrt E
}
e^{-\beta E}
\nonumber\\
&=
\frac{1}{2\sqrt{\pi\beta}}
\exp\!\left(
\frac{\pi^2}{\beta}
\right)
\nonumber\\
&=
2\beta\,
Z_{\mathrm{JT}}^{\mathrm{disk}}(\beta).
\label{eq:source-marked-disk-normalization}
\end{align}
In particular, the factor \(2\beta\) is determined by the exact BF
quantization and is not fixed by imposing the string equation.

At the source-free endpoint,
\begin{equation}
q_*=0,
\qquad
a_*=1,
\label{eq:string-source-free-parameters}
\end{equation}
the insertion carries no curvature source, but its position remains as a
labelled moving point. Forgetting this marking therefore leaves the
remaining sourced BF configuration unchanged. After localization, its
position parametrizes the fiber of the universal curve
\begin{equation}
\pi:\mathcal C_{g,\mathcal A}
\longrightarrow
\overline{\mathcal M}_{g,\mathcal A}.
\label{eq:string-universal-curve}
\end{equation}
Within Hassett geometry, the same fiber may be represented by adjoining an
auxiliary marking of sufficiently small positive stability weight
\(\epsilon\) and taking the limit \(\epsilon\to0^+\) after pushforward. This
auxiliary weight should be distinguished from the BF source parameter,
which is fixed at \(q_*=0\).
The standard pushforward formula for the ordinary or weighted
Weil--Petersson measure along this forgetful map gives the string relation
\cite{Hassett:2003,DoNorbury:2009,Eberhardt:2023rzz}
\begin{equation}
V^{\mathrm{wt}}_{g,n;L+1}
(\mathbf b;\mathbf a,1)
=
\sum_{r=1}^{n}
\int_0^{b_r}
\mathrm db'_r\,b'_r\,
V^{\mathrm{wt}}_{g,n;L}
(b_1,\ldots,b'_r,\ldots,b_n;\mathbf a).
\label{eq:weighted-string-equation}
\end{equation}
The BF construction supplies a direct microscopic interpretation of the
additional marking appearing in this identity: it is precisely the
source-free member \(\mathfrak D^{\mathrm{BF}}_0\) of the same
gauge-invariant defect family.

We now sew the geodesic boundaries to asymptotic JT boundaries. The
Fenchel--Nielsen twist quotient gives the length measure \(b\,db\), whose
derivation and normalization are reviewed in Appendix~\ref{app:BF_gravitational_component}. We define
\begin{align}
Z_{g,n;L}
(\boldsymbol\beta;\mathbf q)
&=
\int_0^\infty
\prod_{s=1}^{n}
\left[
b_s\,\mathrm db_s\,
Z_{\mathrm{tr}}(\beta_s,b_s)
\right]
V^{\mathrm{wt}}_{g,n;L}
(\mathbf b;\mathbf a),
\label{eq:weighted-amplitude-trumpet-gluing}
\end{align}
where
\begin{equation}
Z_{\mathrm{tr}}(\beta,b)
=
\frac{1}{2\sqrt{\pi\beta}}
\exp\!\left(
-\frac{b^2}{4\beta}
\right).
\label{eq:string-trumpet-amplitude}
\end{equation}
The amplitude with one source-free insertion is
\begin{align}
Z^{(0)}_{g,n;L}
(\boldsymbol\beta;\mathbf q)
&=
\int_0^\infty
\prod_{s=1}^{n}
\left[
b_s\,\mathrm db_s\,
Z_{\mathrm{tr}}(\beta_s,b_s)
\right]
V^{\mathrm{wt}}_{g,n;L+1}
(\mathbf b;\mathbf a,1).
\label{eq:source-free-amplitude-trumpet-gluing}
\end{align}

Substituting \eqref{eq:weighted-string-equation} and exchanging the two
length integrations in the \(r\)th term gives
\begin{equation}
\int_0^\infty
b'_r\,\mathrm db'_r\,
F_r(b'_r)
\int_{b'_r}^{\infty}
b_r\,\mathrm db_r\,
Z_{\mathrm{tr}}(\beta_r,b_r),
\label{eq:string-exchanged-length-integral}
\end{equation}
where \(F_r\) contains the remaining moduli-space and boundary integrations.
The trumpet kernel obeys
\begin{align}
\int_{b'}^\infty
b\,\mathrm db\,
Z_{\mathrm{tr}}(\beta,b)
&=
\frac{1}{2\sqrt{\pi\beta}}
\int_{b'}^\infty
b\,\mathrm db\,
\exp\!\left(
-\frac{b^2}{4\beta}
\right)
\nonumber\\
&=
2\beta\,
Z_{\mathrm{tr}}(\beta,b').
\label{eq:string-trumpet-identity}
\end{align}
The \(r\)th boundary channel therefore contributes
\begin{equation}
Z^{(0,r)}_{g,n;L}
(\boldsymbol\beta;\mathbf q)
=
2\beta_r\,
Z_{g,n;L}
(\boldsymbol\beta;\mathbf q).
\label{eq:string-single-boundary-contribution}
\end{equation}
Summing over the \(n\) asymptotic boundaries yields
\begin{equation}
Z^{(0)}_{g,n;L}
(\boldsymbol\beta;\mathbf q)
=
2\left(
\sum_{r=1}^{n}\beta_r
\right)
Z_{g,n;L}
(\boldsymbol\beta;\mathbf q).
\label{eq:BF-string-equation-partition-function}
\end{equation}
Equivalently,
\begin{align}
&
\left\langle
\mathfrak D^{\mathrm{BF}}_0
\prod_{i=1}^{L}
\mathfrak D^{\mathrm{BF}}_{q_i}
\right\rangle_{g,n;\boldsymbol\beta}
\nonumber\\
&\qquad=
2\left(
\sum_{r=1}^{n}\beta_r
\right)
\left\langle
\prod_{i=1}^{L}
\mathfrak D^{\mathrm{BF}}_{q_i}
\right\rangle_{g,n;\boldsymbol\beta}.
\label{eq:BF-string-equation}
\end{align}
For the one-boundary disk, this agrees with the independent BF result
\eqref{eq:source-marked-disk-normalization}.

It is useful to ask whether
\eqref{eq:BF-string-equation} admits a derivation that does not use the
known forgetful-map formula \eqref{eq:weighted-string-equation}. The BF
sewing picture suggests such an interpretation. Since
\(\mathfrak D^{\mathrm{BF}}_0\) carries no curvature source, the insertion
does not change the local holonomy data of the remaining BF configuration.
One may therefore expect its integrated position mode to decompose into
channels in which the source-free marking is moved toward the asymptotic
boundaries. In the \(r\)th channel, the elementary \(q=0\) kernel and BF
sewing suggest the replacement $\mathfrak D^{\mathrm{BF}}_0
\quad\rightsquigarrow\quad
2\beta_r $ on the corresponding boundary state. Summing over the possible boundary
channels then reproduces the right-hand side of
\eqref{eq:BF-string-equation}.

The sewing picture gives a boundary-channel interpretation of the string
relation, but the integral over the compactified position modulus is
evaluated here using the ordinary or weighted forgetful-map identity. In a
blunt chamber, this identity also accounts for degeneration strata on which
the source-free marking meets an admissible defect cluster. Deriving these
boundary and collision terms directly from a BF Ward identity remains an
open problem, as does the extension to the higher Virasoro constraints.

\subsubsection{The dilaton equation}
\label{subsubsec:dilaton-equation}

The dilaton operator is obtained by varying the BF defect at the
source-free endpoint.  Writing \(q_*=1-a_*\), we define
\begin{equation}
\widehat{\mathcal O}_{\mathrm{dil}}
=
\frac{1}{2\pi^2}
\left.
\frac{\partial}{\partial a_*}
\mathfrak D^{\mathrm{BF}}_{1-a_*}
\right|_{a_*=1}.
\label{eq:normalized-BF-dilaton-operator}
\end{equation}
Using the definition of the defect observable gives
\begin{equation}
\widehat{\mathcal O}_{\mathrm{dil}}
=
-\frac{i}{\pi}
\int_\Sigma
\Omega_n(A)\,
\langle n,X\rangle.
\label{eq:BF-dilaton-operator}
\end{equation}
On the gravitational integration cycle,
\(\langle n,X\rangle=i\Phi\) and
\(\Omega_n(A)=\sqrt g\,\mathrm d^2x\), so that
\begin{equation}
\left.
\widehat{\mathcal O}_{\mathrm{dil}}
\right|_{\Gamma_{\mathrm{grav}}}
=
\frac{1}{\pi}
\int_\Sigma
\sqrt g\,\mathrm d^2x\,\Phi.
\label{eq:BF-dilaton-gravitational}
\end{equation}
At \(a_*=1\) the localized curvature source vanishes, but the integrated
operator remains a smooth moving marked point.  The first variation at
this endpoint is therefore the BF realization of the dilaton insertion.

Its normalization is fixed by the exact one-defect disk amplitude,
\begin{equation}
Z_{1\text{-}\mathrm{def}}(\beta;a)
=
\frac{1}{2\sqrt{\pi\beta}}
\exp\!\left(
\frac{\pi^2a^2}{\beta}
\right).
\label{eq:defect-amplitude-opening-fraction}
\end{equation}
Indeed,
\begin{equation}
\frac{1}{2\pi^2}
\left.
\frac{\partial}{\partial a}
Z_{1\text{-}\mathrm{def}}(\beta;a)
\right|_{a=1}
=
2Z_{\mathrm{JT}}^{\mathrm{disk}}(\beta).
\label{eq:disk-dilaton-normalization}
\end{equation}
Thus the BF kernel fixes the factor of \(2\) appearing in the dilaton
equation.

The remaining topological factor is the weighted Euler characteristic of
the localized cone surface.  For a genus-\(g\) surface with \(n\)
geodesic boundaries and cone opening angles
\(\theta_i=2\pi(1-q_i)\), the conical Gauss--Bonnet theorem gives
\begin{equation}
\int_{\Sigma_{\mathrm{reg}}}K\,\mathrm dA
+
2\pi\sum_{i=1}^{L}q_i
=
2\pi(2-2g-n),
\end{equation}
and hence
\begin{equation}
\chi_{\mathbf q}
\equiv
\frac{1}{2\pi}
\int_{\Sigma_{\mathrm{reg}}}K\,\mathrm dA
=
2-2g-n-\sum_{i=1}^{L}q_i.
\label{eq:weighted-Euler-characteristic}
\end{equation}
From the BF viewpoint, this is the natural topological charge entering
the dilaton relation: \(X\) is the multiplier conjugate to the curvature
constraint, while \(\widehat{\mathcal O}_{\mathrm{dil}}\) is its
source-free first variation.

The resulting dilaton equation is
\begin{align}
&
\left\langle
\widehat{\mathcal O}_{\mathrm{dil}}
\prod_{i=1}^{L}
\mathfrak D^{\mathrm{BF}}_{q_i}
\right\rangle_{g,n;\mathbf b}
\nonumber\\
&\qquad=
2\chi_{\mathbf q}
\left\langle
\prod_{i=1}^{L}
\mathfrak D^{\mathrm{BF}}_{q_i}
\right\rangle_{g,n;\mathbf b}.
\label{eq:BF-dilaton-equation}
\end{align}
For \(g=0\), \(n=1\), and \(L=0\),
\(\chi_{\mathbf q}=1\), and
\eqref{eq:BF-dilaton-equation} reduces to the exact disk identity
\eqref{eq:disk-dilaton-normalization}.

Globally, the additional moving marking must be integrated over the
compactified position moduli.  The required statement is the weighted
dilaton pushforward identity,
\begin{equation}
\left.
\frac{1}{2\pi^2}
\frac{\partial}{\partial a_*}
V^{\mathrm{wt}}_{g,n;L+1}
(\mathbf b;\mathbf a,a_*)
\right|_{a_*=1}
=
2\chi_{\mathbf q}\,
V^{\mathrm{wt}}_{g,n;L}
(\mathbf b;\mathbf a),
\label{eq:weighted-dilaton-equation}
\end{equation}
with the appropriate compactification understood in each chamber
\cite{Hassett:2003,Eberhardt:2023rzz}.
Together with the BF identification of the local dilaton operator and
its disk normalization, this gives
\eqref{eq:BF-dilaton-equation}.  Thus the BF description determines the
local operator and its normalization, while the weighted pushforward
identity supplies the global dependence on the compactified position
moduli.  A purely BF derivation of the dilaton equation would therefore
have to reproduce this global pushforward structure, rather than follow from the local BF Ward identities alone.

\section{Genus-zero partition functions from BF data and weighted geometry}
\label{sec:partition}

We now combine the BF defect observable and its exact elliptic kernel with
the compactified moduli-space description of the preceding section to
obtain fixed-order genus-zero gravitational amplitudes.

For several defects, the integrated amplitude depends not only on the
elementary BF kernel but also on the compactification of their
relative-position moduli. In the open sharp chamber, ordinary
Weil--Petersson sewing is sufficient. In blunt chambers, admissible
collisions produce additional wall-crossing contact terms
\cite{Eberhardt:2023rzz,Lin:2023trc,Kruthoff:2024rpx}.

We use the opening fractions and fractional deficits introduced in
\eqref{eq:descendant-angle-parameters},
\begin{equation}
a_i
=
\frac{\theta_i}{2\pi},
\qquad
q_i
=
1-a_i.
\label{eq:section5-parameters}
\end{equation}
For a cluster \(I\) of aligned elliptic sources, the BF
source-composition rule gives
\begin{equation}
q_I
=
\sum_{i\in I}q_i,
\qquad
a_I
=
1-q_I.
\label{eq:section5-cluster-parameter}
\end{equation}
As shown in Section~\ref{subsec:admissible}, the fused source remains in
the gravitational cone sector when \(q_I\leq1\), with equality giving the
cusp limit.

We call \(q_i>\tfrac12\) the open sharp regime and
\(0<q_i<\tfrac12\) the blunt regime. The chamber walls occur at
\(\sum_{i\in I}q_i=1\). Crossing a wall does not change the elementary BF
operator; it changes which collision strata are included in the
compactified position space.

Throughout this section, we work on \(\Gamma_{\mathrm{grav}}\) and at
fixed order in the defect fugacities. All disk amplitudes are written with
the overall topological factor \(e^{S_0}\) suppressed. We evaluate the
weighted amplitudes explicitly only at genus zero and do not address a
nonperturbative completion of the genus expansion.

\subsection{Sharp defects}
\label{subsec:sharp}

In the open sharp chamber, every pair satisfies
\begin{equation}
q_i+q_j
>
1.
\label{eq:section5-sharp-pair-condition}
\end{equation}
No two defect markings can therefore coincide while remaining in the
gravitational cone sector. The relevant compactification is the ordinary
Deligne--Mumford one, and the cone-surface volumes are obtained by analytic
continuation of the usual Weil--Petersson geometry
\cite{Mirzakhani:2006,DoNorbury:2009}. The boundary
\(q_i=q_j=\tfrac12\) corresponds to a cuspidal fused source and can be
reached by continuity
\cite{Eberhardt:2023rzz,Anagnostou:2022cmq,Anagnostou:2023jiy}.

\subsubsection{Elementary defect input}
\label{subsubsec:elementary-defect-input}

We first recall the one-defect result of
Section~\ref{sec:quantization}. In our spectral and sewing convention, a
defect of opening fraction \(a\) contributes
\begin{equation}
Z_{1\text{-}\mathrm{def}}(\beta;a)
=
\frac{1}{2\sqrt{\pi\beta}}
\exp\!\left(
\frac{\pi^2a^2}{\beta}
\right).
\label{eq:section5-one-defect-amplitude}
\end{equation}
This is the elementary BF input for the multi-defect amplitudes. The
dependence on the relative defect positions enters separately through the
compactified moduli-space integral.

\subsubsection{Stable amplitudes in the sharp chamber}
\label{subsubsec:sharp-multidefect}

Since the defect markings remain distinct, no additional collision strata
or contact vertices are required. Once the elementary BF defect amplitude
has been fixed, all stable multi-defect amplitudes follow from ordinary
Weil--Petersson sewing.

For a disk with \(m\geq2\) defects, removing the asymptotic trumpet leaves
a stable genus-zero core with one geodesic boundary. Its contribution is
\begin{align}
Z_{0,1}^{(m)}(\beta)
&=
\frac{1}{m!}
\sum_{s_1,\ldots,s_m}
\lambda_{s_1}\cdots\lambda_{s_m}
\int_0^\infty
b\,db\,
Z_{\mathrm{tr}}(\beta,b)
\nonumber\\
&\qquad\times
V_{0,m+1}
\left(
b,
2\pi i a_{s_1},
\ldots,
2\pi i a_{s_m}
\right).
\label{eq:section5-sharp-WP-gluing}
\end{align}
Here
\begin{equation}
Z_{\mathrm{tr}}(\beta,b)
=
\frac{1}{2\sqrt{\pi\beta}}
\exp\!\left(
-\frac{b^2}{4\beta}
\right),
\label{eq:section5-trumpet-amplitude}
\end{equation}
and the cone points are implemented by the standard continuation
\(b_i=2\pi i a_{s_i}\)
\cite{DoNorbury:2009,Witten:2020wvy,Maxfield:2020ale}.
The one-defect disk is unstable and is instead supplied by
\eqref{eq:section5-one-defect-amplitude}.

As a normalization check, \(V_{0,3}=1\) gives
\begin{equation}
Z_{0,1}^{(2)}(\beta)
=
\frac{\sqrt{\beta}}{2\sqrt{\pi}}
\left(
\sum_s\lambda_s
\right)^2.
\label{eq:section5-second-order}
\end{equation}

At genus zero, the same construction applies to any stable number of
asymptotic boundaries: the compact core is governed by \(V_{0,n+m}\), with
a trumpet sewn to each asymptotic boundary. Thus, in the sharp chamber,
the BF construction introduces no multi-defect contact data beyond the
elementary insertion. All higher orders in the defect fugacities are
generated by ordinary Weil--Petersson sewing.

\subsection{Blunt defects}
\label{subsec:blunt}

We now turn to chambers in which weighted coincidence strata are present.
In contrast to the open sharp chamber, subsets \(I\) satisfying
\(q_I\leq1\) may collide. The corresponding change of compactification
produces contact corrections that are not contained in the ordinary
resolved Weil--Petersson integral alone.

\subsubsection{Multi-defect contact terms}
\label{subsubsec:cluster-amplitudes}

We relate BF source fusion to the multi-defect contact terms of the
deformed-JT string equation. The wall-crossing formula is known
\cite{Turiaci:2020fjj,Eberhardt:2023rzz}. Here the BF source equation
determines the fused parameter, while restriction to
\(\Gamma_{\mathrm{grav}}\) determines when the fused source is admissible.

Consider \(L\geq2\) identical defects of fractional deficit \(q\) and
fugacity \(\lambda\). By \eqref{eq:section5-cluster-parameter}, their
fused opening fraction is \(a_L=1-Lq\), and the cluster is admissible when
\(a_L\geq0\), or equivalently \(Lq\leq1\). Define
\begin{equation}
z_L
=
\frac{\pi^2a_L^2}{\beta},
\qquad
\mathcal N_L(\beta)
=
\frac{\lambda^L}{L!}
\frac{2^{L-2}\beta^{L-\frac32}}{\sqrt{\pi}}.
\label{eq:section5-cluster-abbreviations}
\end{equation}

Treating the coincident defects as the fused elliptic source, the disk
fixed-point calculation gives
\begin{equation}
Z_{\mathrm{coinc}}^{(L)}(\beta)
=
\mathcal N_L(\beta)e^{z_L}.
\label{eq:section5-coincident-amplitude}
\end{equation}
This is the complete local fixed-point block; no subtraction enters the
localization calculation.

To incorporate this block into the compactified moduli-space description,
we compare it with the resolved side of the collision wall. As derived in
Appendix~\ref{app:general-L-wall-crossing}, the resolved rational tail
contributes
\begin{equation}
Z_{\mathrm{res}}^{(L)}(\beta)
=
\mathcal N_L(\beta)
\sum_{r=0}^{L-2}
\frac{z_L^r}{r!}.
\label{eq:section5-resolved-amplitude}
\end{equation}
The sum terminates at \(r=L-2\), reflecting
\(\dim_{\mathbb C}\overline{\mathcal M}_{0,L+1}=L-2\).

Taking the resolved moduli-space integral as the baseline, wall crossing
replaces this rational-tail contribution by the coincident fixed-point
block. The resulting contact correction is
\begin{equation}
Z_{\mathrm{disk,contact}}^{(L)}(\beta)
=
\mathcal N_L(\beta)
\left[
e^{z_L}
-
\sum_{r=0}^{L-2}
\frac{z_L^r}{r!}
\right],
\qquad
a_L\geq0.
\label{eq:section5-ZL}
\end{equation}

Under the standard genus-zero disk transform,
\eqref{eq:section5-ZL} reproduces the known \(L\)-defect contact term in
the deformed-JT string function
\cite{Turiaci:2020fjj,Eberhardt:2023rzz}. The local BF source equation
therefore fixes the fused elliptic parameter, while
\(\Gamma_{\mathrm{grav}}\) selects the admissible clusters and the global
wall-crossing geometry supplies their contact corrections.

\subsubsection{Genus-zero cluster expansion}
\label{subsubsec:genus-zero-cluster-expansion}

For identical defects of fractional deficit \(q>0\), the condition
\(Lq\leq1\) implies that the cluster expansion terminates at
\begin{equation}
L_{\max}
=
\left\lfloor
\frac{1}{q}
\right\rfloor.
\label{eq:section5-Lmax}
\end{equation}
At the endpoint \(Lq=1\), the fused source is cuspidal. For \(L\geq2\),
the corresponding contact term vanishes there, so including the endpoint
does not change the string function.

Combining the elementary defect insertion with the contact corrections
above gives
\begin{align}
\mathcal F(u;\lambda,q)
&=
\frac{\sqrt{u}}{2\pi}
I_1(2\pi\sqrt{u})
\nonumber\\
&\quad+
\sum_{L=1}^{L_{\max}}
\frac{\lambda^L}{L!}
\left(
\frac{2\pi(1-Lq)}{\sqrt{u}}
\right)^{L-1}
I_{L-1}
\!\left(
2\pi(1-Lq)\sqrt{u}
\right).
\label{eq:section5-full-genus-zero-string-function}
\end{align}
The \(L=1\) term is the elementary contribution
\(\lambda I_0(2\pi(1-q)\sqrt{u})\), while the terms with \(L\geq2\) are
the admissible multi-defect contact vertices.

Equation~\eqref{eq:section5-full-genus-zero-string-function} is the known
genus-zero deformed-JT string function
\cite{Turiaci:2020fjj,Eberhardt:2023rzz}. The BF description gives a
simple interpretation of its chamber structure. The wall
\begin{equation}
q
=
\frac{1}{L}
\label{eq:section5-identical-cluster-wall}
\end{equation}
is precisely where the fused \(L\)-defect source reaches the cuspidal
limit \(a_L=0\). For \(q<1/L\), this source belongs to the gravitational
cone sector. For \(q>1/L\), it leaves that sector, and the corresponding
contact vertex is no longer present.

\section{Discussion and outlook}
\label{sec:discussion}

We have constructed a gauge-invariant BF representative of a conical defect
relative to the elliptic reduction that defines the gravitational sector.
On \(\Gamma_{\mathrm{grav}}\), this observable reproduces the metric defect
insertion, its distributional curvature source, and its elliptic monodromy.
The elliptic reduction is fixed background data rather than a field
integrated over in the path integral. Here, gauge invariance refers to
simultaneous transformations of \(A\), \(X\), and the chosen section \(n\).
For a single defect on the disk, quotienting by the position orbit reduces the observable to a fixed elliptic sector. The exact BF/Schwarzian
quantization of this sector provides the elementary kernel and disk
amplitude used throughout the gravitational calculation.

The same local observable describes both sharp and blunt defects. Their
difference appears only when the positions of several defects are
integrated. For aligned sources, the BF source equation gives
\begin{equation}
q_I
=
\sum_{i\in I}q_i,
\end{equation}
while restriction to the compactified gravitational cone sector requires
\begin{equation}
q_I\leq1.
\end{equation}
This is precisely the coincidence condition for marked points with Hassett
weights \(q_i\)~\cite{Hassett:2003}. In the open sharp chamber, the defect
markings cannot coincide. In the blunt regime, by contrast, additional
collision loci are allowed whenever the total weight of the colliding
subset does not exceed one. The corresponding amplitudes therefore involve different compactifications of the relative-position moduli.

The BF description also accounts for part of the tautological geometry of
these moduli spaces. After quotienting by the position orbit of a moving
defect, its compact rotational stabilizer defines the circle bundle
associated with the cotangent line at the marked point. The first Chern
class of this bundle is \(\psi_i\). The BF source equation further
determines the label of a fused cluster and the subsets for which a
collision is admissible. However, completing these local statements into a
compact moduli problem requires the global geometry of weighted stable
curves. In particular, the reduction morphisms and the coefficients of the
conical Weil--Petersson contact terms are provided by the known geometric
results~\cite{Anagnostou:2022cmq,Anagnostou:2023jiy,
Eberhardt:2023rzz}; they do not follow from the local sourced-flatness
equation alone.

Combining the BF description with the weighted Weil--Petersson class, we
obtain the fixed-order genus-zero amplitudes in the sharp and blunt
chambers, including the contact contributions from admissible defect
collisions. Where matrix-model expressions are available, the elementary disk amplitude and the genus-zero cluster coefficients agree with the
corresponding terms in deformed JT
gravity~\cite{Witten:2020wvy,Maxfield:2020ale,Turiaci:2020fjj}. Thus, the
matrix model serves as a comparison rather than as an input in the
definition of the BF observable, its elementary kernel, or its
source-composition rule.

Several qualifications remain. The BF action alone does not define
Euclidean JT gravity: the integration cycle \(\Gamma_{\mathrm{grav}}\)
selects the gravitational configurations and the appropriate contour for
the BF scalar. A global construction of this cycle in the complexified
field space remains open. Moreover, we have evaluated the weighted
amplitudes explicitly only at genus zero. Their higher-genus extension
requires control of weighted boundary classes, nonseparating degenerations,
and sewing relations. The forgetful-map and pushforward identities used
here are additional geometric input. Deriving them directly from a BF or
BRST framework, and organizing the resulting structure as a weighted
topological gravity, are interesting questions for future work.

\section*{Acknowledgements}
We thank E.~Witten for considering our question and sharing his thoughts,
and in particular for pointing us to several relevant references.
We thank OpenAI's ChatGPT (GPT-5.6 Sol) for useful discussions and feedback, particularly for suggesting Hassett geometry as a relevant framework, as well as for assistance with the exploration of the literature and the writing of the manuscript. The research of W.G. is supported by the National Natural Science Foundation
of China (NSFC) under Grant No.~12575077.

\appendix

\section{The BF measure on the smooth gravitational localization locus}
\label{app:moment_map_reduction}

In this appendix, we explain in more detail why the BF path integral,
when restricted to the gravitational sector, induces the
Weil--Petersson measure on the moduli space of hyperbolic surfaces. There
are two logically distinct issues. The first is local. For the noncompact
gauge group \(\mathrm{PSL}(2,\mathbb R)\), the invariant bilinear form on the Lie
algebra is indefinite and therefore cannot be used directly to define a
positive gauge-fixing measure. The second issue is global. The moduli space
of flat \(\mathrm{PSL}(2,\mathbb R)\) connections has several connected components,
only one of which describes the hyperbolic geometries relevant to JT
gravity.

For a compact gauge group, the relation between two-dimensional gauge
theory, localization, and the symplectic measure on the moduli space of
flat connections was developed by Witten and by Blau and
Thompson~\cite{Witten:1992xu,Blau:1993hj}. For a noncompact group, the
same argument requires an additional prescription because the invariant
bilinear form is indefinite. This issue was discussed in the JT context
in~\cite{Saad:2019lba}. As explained there, following a suggestion
of Witten and the perturbative treatment of noncompact Chern--Simons
theory in~\cite{Bar-Natan:1991fix}, one expects the corresponding
local argument to extend to the noncompact BF theory relevant here.

The use of a positive but non-invariant auxiliary metric for the
noncompact theory was already explained in the JT context in~\cite{Saad:2019lba}. We adopt the same prescription here, using the
auxiliary metric only to define the local gauge-fixed measure around a
regular irreducible flat connection. We then spell out the corresponding
BF gauge-fixing calculation and the cancellation of the nonzero-mode
determinants. Although the main ingredients may be familiar to experts, we
have not found this calculation written explicitly for the present BF
setup. We include it for completeness and to state clearly the assumptions
entering the measure used in the main text.

The local calculation does not select a gravitational component. This
additional global input is supplied by the integration prescription
\(\Gamma_{\mathrm{grav}}\). Our argument also applies only to the smooth
locus of separated cone points. It does not construct the weighted
compactification or determine the contact terms supported on collision
strata.

\subsection{Local gauge fixing and the reduced measure}
\label{app:local_BF_measure}

Let \(\mathcal A(\Sigma)\) be the affine space of connections on an
oriented surface \(\Sigma\). The invariant bilinear form defines the
Atiyah--Bott symplectic form
\begin{equation}
\Omega_{\mathcal A}(a,b)
=
\int_\Sigma
\langle a\wedge b\rangle.
\label{eq:app-AB-form}
\end{equation}
For a gauge transformation generated by an adjoint-valued zero-form
\(\epsilon\), with \(\epsilon|_{\partial\Sigma}=0\), one has
\begin{equation}
V_\epsilon(A)
=
d_A\epsilon.
\label{eq:app-gauge-vector}
\end{equation}
The corresponding Hamiltonian is
\begin{equation}
H_\epsilon(A)
=
\int_\Sigma
\langle\epsilon,F_A\rangle.
\label{eq:app-gauge-Hamiltonian}
\end{equation}
Integration by parts gives
\begin{equation}
\iota_{V_\epsilon}\Omega_{\mathcal A}
=
-\delta H_\epsilon.
\label{eq:app-moment-map-identity}
\end{equation}
Thus the gauge action is Hamiltonian, with the curvature, $F_A$, as its
moment map, in the convention
$\iota_{V_\epsilon}\Omega_{\mathcal A}
=-\delta\langle\mu,\epsilon\rangle$.
In particular, the zero level of the moment map is the space of flat
connections.

To define a positive gauge-fixed measure, choose a Cartan involution \(T\)
and introduce, for adjoint-valued forms of the same degree,
\begin{equation}
(\alpha,\beta)_T
\equiv
\int_\Sigma
\langle\alpha\wedge *T\beta\rangle.
\label{eq:app-T-inner-product}
\end{equation}
We choose the sign of the invariant pairing so that
\((\cdot,\cdot)_T\) is positive. On one-forms, define
\begin{equation}
J
=
*T,
\qquad
g_T(a,b)
=
(a,b)_T
=
\Omega_{\mathcal A}(a,Jb).
\label{eq:app-compatible-metric}
\end{equation}
Since \(T^2=1\) and \(*^2=-1\) on one-forms,
\begin{equation}
J^2
=
-1.
\label{eq:app-J-square}
\end{equation}
The metric \(g_T\) is an auxiliary gauge-fixing structure. It need not be
invariant under the full noncompact gauge group, while
\(\Omega_{\mathcal A}\) remains unchanged.

We expand around a regular irreducible flat connection \(A\) and write
\(D=d_A\). The adjoint \(D^{\dagger,T}\) is defined by
\begin{equation}
(D\epsilon,a)_T
=
(\epsilon,D^{\dagger,T}a)_T.
\label{eq:app-T-adjoint}
\end{equation}
The tangent space to the gauge orbit is
\(\mathcal V_A=\operatorname{im}D\). On the flat locus, the gauge orbit is
isotropic:
\begin{equation}
\Omega_{\mathcal A}(D\epsilon,D\eta)
=
0.
\label{eq:app-orbit-isotropic}
\end{equation}
Equations~\eqref{eq:app-moment-map-identity} and
\eqref{eq:app-compatible-metric} then identify
\(J\mathcal V_A\) with the directions normal to the flatness constraint.
Locally, a connection fluctuation can therefore be decomposed as
\begin{equation}
a
=
a_0+D\epsilon+JD\eta,
\label{eq:app-fluctuation-decomposition}
\end{equation}
where
\begin{equation}
a_0
\in
\mathcal H_A
\equiv
\bigl(
\mathcal V_A\oplus J\mathcal V_A
\bigr)^{\perp_{g_T}}.
\label{eq:app-reduced-tangent-space}
\end{equation}
The space \(\mathcal H_A\) represents the tangent space to the reduced
moduli space. With fixed boundary conjugacy classes, it is represented by
the corresponding parabolic cohomology.

We now show that the nonzero-mode determinants cancel. Define
\begin{equation}
M_A
=
D^{\dagger,T}D
\label{eq:app-M-operator}
\end{equation}
on adjoint-valued zero-forms. Let \(u_\lambda\) be an orthonormal nonzero
eigenmode:
\begin{equation}
M_Au_\lambda
=
\lambda u_\lambda,
\qquad
\lambda>0,
\qquad
(u_\lambda,u_\lambda)_T=1.
\label{eq:app-M-eigenmode}
\end{equation}
It follows from \eqref{eq:app-T-adjoint} that
\begin{equation}
\lVert Du_\lambda\rVert_T^2
=
\lVert JDu_\lambda\rVert_T^2
=
\lambda.
\label{eq:app-mode-norms}
\end{equation}
We expand the gauge and normal parameters as
\begin{equation}
\epsilon
=
\sum_{\lambda>0}
\epsilon_\lambda u_\lambda,
\qquad
\eta
=
\sum_{\lambda>0}
\eta_\lambda u_\lambda.
\label{eq:app-mode-expansion}
\end{equation}
For each eigenvalue \(\lambda\), the induced metric on the corresponding
orbit and normal directions is
\begin{equation}
ds_\lambda^2
=
\lambda\,
(d\epsilon_\lambda)^2
+
\lambda\,
(d\eta_\lambda)^2.
\label{eq:app-single-mode-metric}
\end{equation}
The associated volume element is therefore
\begin{equation}
d\mu_\lambda
=
\lambda\,
d\epsilon_\lambda\,d\eta_\lambda.
\label{eq:app-single-mode-measure}
\end{equation}
Taking the product over the nonzero eigenvalues gives
\begin{align}
\mathcal D A
&=
\mathcal D a_0\,
\mathcal D(D\epsilon)\,
\mathcal D(JD\eta)
\nonumber\\
&=
\left(
\prod_{\lambda>0}\lambda
\right)
\mathcal D a_0\,
\mathcal D\epsilon\,
\mathcal D\eta
\nonumber\\
&=
\det\nolimits' M_A\,
\mathcal D a_0\,
\mathcal D\epsilon\,
\mathcal D\eta,
\label{eq:app-measure-Jacobian}
\end{align}
up to a field-independent normalization. 

The BF scalar imposes the moment-map constraint:
\begin{equation}
\int_{\Gamma_X}
\mathcal D X\,
\exp\left[
-i\int_\Sigma
\langle X,F_A\rangle
\right]
\propto
\delta_{\Gamma_X}(F_A),
\label{eq:app-flatness-delta}
\end{equation}
where the contour \(\Gamma_X\) is inherited from the gravitational
integration prescription, whose role will be discussed in
Section~\ref{app:BF_gravitational_component}.

We next impose the flatness condition and remove the gauge-orbit
directions. On the flat background,
\begin{equation}
D_Aa_0
=
0,
\qquad
D_A^2
=
0.
\label{eq:app-closed-zero-orbit-modes}
\end{equation}
The linearized curvature produced by the normal fluctuation is
\begin{equation}
\delta F_A
=
D_AJD_A\eta.
\label{eq:app-linearized-curvature-normal-mode}
\end{equation}
This defines a map from adjoint-valued zero-forms to adjoint-valued
two-forms. To identify this map, we pair it with an arbitrary gauge
parameter \(\epsilon\). Using the moment-map identity and the compatibility
of \(g_T\) with \(\Omega_{\mathcal A}\), we obtain
\begin{align}
\int_\Sigma
\left\langle
\epsilon,
D_AJD_A\eta
\right\rangle
&=
\delta H_\epsilon(JD_A\eta)
\nonumber\\
&=
-(\epsilon,M_A\eta)_T.
\label{eq:app-normal-curvature-pairing}
\end{align}
Thus, after using the fixed \(T\)-metric and the Hodge star to identify
adjoint-valued two-forms with zero-forms, the linearized curvature map is
represented by \(-M_A\). The overall sign does not affect the functional
Jacobian, and the flatness delta function therefore gives
\begin{equation}
\delta_{\mathrm{fun}}(M_A\eta)
=
\bigl(
\det\nolimits' M_A
\bigr)^{-1}
\delta_{\mathrm{fun}}(\eta).
\label{eq:app-flatness-Jacobian}
\end{equation} 

The orbit directions are removed using the \(T\)-modified Lorenz gauge
\begin{equation}
D_A^{\dagger,T}a
=
0.
\label{eq:app-Lorenz-gauge}
\end{equation}
Under a gauge displacement
\begin{equation}
a
\longmapsto
a+D_A\epsilon,
\label{eq:app-linearized-gauge-displacement}
\end{equation}
the gauge-fixing function varies as
\begin{equation}
D_A^{\dagger,T}D_A\epsilon
=
M_A\epsilon.
\label{eq:app-gauge-fixing-variation}
\end{equation}
Hence the gauge-fixing delta function gives
\begin{equation}
\delta_{\mathrm{fun}}(M_A\epsilon)
=
\bigl(
\det\nolimits' M_A
\bigr)^{-1}
\delta_{\mathrm{fun}}(\epsilon),
\label{eq:app-gauge-fixing-Jacobian}
\end{equation}
while the corresponding Faddeev--Popov determinant is
\begin{equation}
\Delta_{\mathrm{FP}}
=
\det\nolimits' M_A.
\label{eq:app-FP-determinant}
\end{equation}

Combining these
factors with \eqref{eq:app-measure-Jacobian} and
\eqref{eq:app-flatness-Jacobian}, we obtain
\begin{equation}
\det\nolimits' M_A\,
\bigl(
\det\nolimits' M_A
\bigr)^{-1}
\bigl(
\det\nolimits' M_A
\bigr)^{-1}
\det\nolimits' M_A
=
1.
\label{eq:app-determinant-cancellation}
\end{equation}
Thus, no moduli-dependent determinant remains from the nonzero modes. This
statement is understood on the regular irreducible locus and with a common
\(T\)-compatible regularization.

It remains to determine the measure of the zero modes. Let \(m^I\) be
local coordinates on the smooth reduced space and let
\(\alpha_I\in\mathcal H_A\) represent the corresponding tangent vectors.
The induced metric and symplectic form are
\begin{equation}
G_{IJ}
=
g_T(\alpha_I,\alpha_J),
\qquad
\omega_{IJ}
=
\int_\Sigma
\langle\alpha_I\wedge\alpha_J\rangle.
\label{eq:app-zero-mode-forms}
\end{equation}
The compatibility relation in \eqref{eq:app-compatible-metric} descends to
the reduced tangent space. In a \(J\)-adapted orthonormal basis, the
zero-mode measure is therefore
\begin{equation}
\sqrt{\det G}\,
d^{2d}m
=
\operatorname{Pf}(\omega_{IJ})\,
d^{2d}m
=
\frac{\omega_{\mathrm{red}}^d}{d!}.
\label{eq:app-Liouville-measure}
\end{equation}
The reduced form \(\omega_{\mathrm{red}}\) is the
Atiyah--Bott--Goldman symplectic form. Hence the local gauge-fixed BF path
integral induces the Liouville measure used in Section~\ref{subsec:correlation-functions}, up to a field-independent
normalization.

Because the Cartan metric is only an auxiliary gauge-fixing structure and is not invariant under the full noncompact gauge group, the preceding calculation should be understood locally on a chosen slice through the regular irreducible locus.  Its conclusion
is the cancellation of the moduli-dependent nonzero-mode factors within a common regularization scheme; it is not a construction of the global noncompact BF integration cycle.

\subsection{The gravitational component and sewing normalization}
\label{app:BF_gravitational_component}

The preceding calculation is local on the smooth character variety. It
does not determine which connected component should be included in the
gravitational path integral. The integration prescription
\(\Gamma_{\mathrm{grav}}\) supplies this global input by selecting the
positive maximal-Euler-class component,
\begin{equation}
\mathcal X_{\mathrm{grav}}^+
(\Sigma;\mathbf b)
\subset
\mathcal X_{PSL(2,\mathbb R)}
(\Sigma;\mathcal C_{\mathbf b}).
\label{eq:app-gravitational-component}
\end{equation}
This component is naturally identified with the Teichm\"uller space of
oriented hyperbolic surfaces with the prescribed boundary lengths:
\begin{equation}
\mathcal X_{\mathrm{grav}}^+
(\Sigma;\mathbf b)
\simeq
\mathcal T_{g,n}(\mathbf b).
\label{eq:app-Teichmuller-identification}
\end{equation}

On this component, the holonomy map identifies the Goldman symplectic form
with the Weil--Petersson form. With the invariant pairing and length
convention used in the main text,
\begin{equation}
\left.
\omega_{\mathrm{Goldman}}
\right|_{\mathcal X_{\mathrm{grav}}^+}
=
\omega_{\mathrm{WP}}.
\label{eq:app-Goldman-WP}
\end{equation}
The reduced BF measure is therefore
\begin{equation}
d\mu_{\mathrm{BF}}^{\mathrm{grav,red}}
=
C_{\mathrm{top}}\,
\frac{
\omega_{\mathrm{WP}}^{3g-3+n}
}{
(3g-3+n)!
},
\label{eq:app-BF-WP-measure}
\end{equation}
where \(C_{\mathrm{top}}\) is a field-independent normalization not fixed
by the local determinant calculation.

For fixed elliptic conjugacy classes, the same argument gives the conical
Weil--Petersson measure on the smooth locus of separated cone points. Its
extension to the weighted compactification, including the coefficients of
the collision strata, requires the global Hassett and conical
Weil--Petersson geometry. It does not follow from the local cancellation
\eqref{eq:app-determinant-cancellation}.

Passing from marked hyperbolic structures to gravitational moduli further
requires quotienting by the mapping class group:
\begin{equation}
\mathcal M_{g,n}(\mathbf b)
=
\mathcal T_{g,n}(\mathbf b)/
\operatorname{MCG}(\Sigma),
\label{eq:app-MCG-quotient}
\end{equation}
with the usual orbifold weights.

We fix the remaining normalization by requiring
\begin{equation}
V_{0,3}(L_1,L_2,L_3)
=
1
\label{eq:app-pants-normalization}
\end{equation}
and by normalizing the cylinder to act as the identity. Functorial sewing
then fixes
\begin{equation}
C_{\mathrm{top}}
=
1
\label{eq:app-core-normalization}
\end{equation}
for the dynamical core. A factor depending only on the Euler
characteristic is kept separately as \(e^{S_0\chi(\Sigma)}\).

Finally, let \((b,\tau)\) be the Fenchel--Nielsen coordinates associated
with an internal geodesic. Locally,
\begin{equation}
\omega_{\mathrm{WP}}
=
db\wedge d\tau+\cdots,
\qquad
\tau\sim\tau+b.
\label{eq:app-FN-coordinates}
\end{equation}
Integrating over the twist orbit gives
\begin{equation}
d\mu_{\mathrm{sew}}(b)
=
b\,db.
\label{eq:app-bdb-measure}
\end{equation}
This is the length measure used in the gluing formulas of the main text.

For every stable topology \(2g-2+n>0\), we therefore obtain
\begin{equation}
Z_{g,n}^{\mathrm{BF,grav,core}}(\mathbf b)
=
\int_{\mathcal M_{g,n}(\mathbf b)}
\frac{
\omega_{\mathrm{WP}}^{3g-3+n}
}{
(3g-3+n)!
}
=
V_{g,n}(\mathbf b).
\label{eq:app-BF-WP-volume}
\end{equation}
The disk and cylinder are unstable and are defined separately by the
asymptotic boundary quantization and the BFV identity pairing,
respectively. Equation~\eqref{eq:app-BF-WP-volume} concerns only the
localized dynamical core. It does not include the asymptotic trumpet
factors, the Euler-characteristic weight, or the contact contributions
arising from the compactification of the conical-defect position space.

\section{Resolved contribution of an admissible
\texorpdfstring{\(L\)}{L}-defect collision}
\label{app:general-L-wall-crossing}

This appendix derives the resolved contribution used in
Section~\ref{subsubsec:cluster-amplitudes}. Consider \(L\) identical
defects of opening fraction \(a\) and fractional deficit \(q=1-a\).
When their positions approach one another with aligned elliptic sources,
the BF source-composition rule gives
\begin{equation}
q_L
=
Lq,
\qquad
a_L
=
1-q_L
=
1-L(1-a).
\label{eq:app-fused-parameter}
\end{equation}
We restrict to \(a_L\geq0\), for which the fused source remains in the
compactified gravitational cone sector.

Before the \(L\)-fold collision is contracted, the resolved side of the
wall retains the relative positions of the \(L\) markings on a rational
component with one attaching point \(p_\bullet\). This component is
therefore parametrized by \(\overline{\mathcal M}_{0,L+1}\). Cutting at
the attachment produces a geodesic boundary of length \(b\), and we define
\begin{equation}
B
=
\frac{b}{2\pi}.
\label{eq:app-resolved-boundary-parameter}
\end{equation}

The local BF equation fixes the fused parameter \(a_L\), while the
dependence on the resolved position moduli is governed by the conical
Weil--Petersson class. Using the pullback relations of~\cite{Eberhardt:2023rzz}, its restriction to the resolved component
can be written as
\begin{equation}
\Omega_L
\equiv
\frac{[\omega_{\mathrm{WP}}^{\mathrm{cone}}]}{2\pi^2}
=
B^2\psi_\bullet
+
\kappa_1
-
a^2\sum_{i=1}^{L}\psi_i
+
\sum_{\substack{
I\subsetneq\{1,\ldots,L\}\\
|I|\geq2
}}
a_I^2\delta_{0,I},
\qquad
a_I
=
1-|I|(1-a).
\label{eq:app-WP-class}
\end{equation}
Only proper subsets \(I\) occur in this expression. The full set
corresponds to the \(L\)-fold contraction itself and is not a boundary
divisor retained on the resolved component.

For identical defects, the same pullback relations give the identity
\begin{equation}
\kappa_1
-
a^2\sum_{i=1}^{L}\psi_i
-
a_L^2\psi_\bullet
+
\sum_{\substack{
I\subsetneq\{1,\ldots,L\}\\
|I|\geq2
}}
a_I^2\delta_{0,I}
=
0.
\label{eq:app-rational-tail}
\end{equation}
This is the global geometric input of the calculation; in particular, it
does not follow from the local BF source equation. Combining
\eqref{eq:app-WP-class} and \eqref{eq:app-rational-tail}, we obtain
\begin{equation}
\Omega_L
=
\left(
B^2+a_L^2
\right)
\psi_\bullet.
\label{eq:app-WP-collapse}
\end{equation}
Thus all dependence on the individual markings and their proper
subcollisions is collected into the single fused parameter \(a_L\).

Since
\(\dim_{\mathbb C}\overline{\mathcal M}_{0,L+1}=L-2\) and
\begin{equation}
\int_{\overline{\mathcal M}_{0,L+1}}
\psi_\bullet^{L-2}
=
1
\label{eq:app-rational-tail-psi-integral}
\end{equation}
\cite{Witten:1990hr,Kontsevich:1992ti}, the resolved volume is
\begin{align}
V_{0,L,1}^{\mathrm{res}}(b;a)
&=
\frac{(2\pi^2)^{L-2}}{(L-2)!}
\int_{\overline{\mathcal M}_{0,L+1}}
\Omega_L^{L-2}
\nonumber\\
&=
\frac{
\left[
b^2+(2\pi a_L)^2
\right]^{L-2}
}{
2^{L-2}(L-2)!
}.
\label{eq:app-resolved-volume}
\end{align}
For \(L=2\), this reduces to
\(V_{0,2,1}^{\mathrm{res}}=1\), as expected from
\(\overline{\mathcal M}_{0,3}\).

The result separates the two inputs entering the wall-crossing
calculation. The BF source equation determines the fused elliptic
parameter \(a_L\), while the conical Weil--Petersson pullback relations
determine how the resolved relative-position moduli reduce to the
cotangent-line class at the attaching point.

\end{document}